\documentclass[%
 reprint,
 amsmath,amssymb,
 aps,
 prb,
]{revtex4-2}

\usepackage{ucs}
\usepackage[utf8x]{inputenc}
\usepackage{amssymb}
\usepackage{bbm}
\usepackage[greek,english]{babel}

\usepackage{multirow}
\usepackage{longtable}
\usepackage{ragged2e}
\usepackage{graphicx}
\usepackage{dcolumn}
\usepackage{bm}
\usepackage{hyperref}
\usepackage{stmaryrd}

\newcommand{\sixj}[6]
{
  \left\{ \begin{array}{ccc} #1 & #2 & #3 \\
    #4 & #5 & #6 \end{array} \right\}
}

\begin{document}

\preprint{APS/123-QED}

\title{Transition intensities of trivalent lanthanide ions in solids:\\Revisiting the Judd-Ofelt theory}

\author{Gohar Hovhannesyan}
 \email{gohar\textunderscore hovhannesyan@etu.u-bourgogne.fr}

\author{Vincent Boudon}

\author{Maxence Lepers}
 \email{maxence.lepers@u-bourgogne.fr}

\affiliation{Laboratoire Interdisciplinaire Carnot de Bourgogne, UMR 6303 CNRS-Univ.~Bourgogne Franche-Comt\'{e}, 9 Avenue Alain Savary, BP 47 870, F-21078 Dijon cedex, France}

\date{\today}

\begin{abstract}
We present a modified version of the Judd-Ofelt theory, which describes the intensities of f-f transitions by trivalent lanthanide ions (Ln$^{3+}$) in solids. In our model, the properties of the dopant are calculated with well-established atomic-structure techniques, while the influence of the crystal-field potential is described by three adjustable parameters. By applying our model to europium (Eu$^{3+}$), well-known to challenge the standard Judd-Ofelt theory, we are able to give a physical insight into all the transitions within the ground electronic configuration, and also to reproduce quantitatively experimental absorption oscillator strengths. Our model opens the possibility to interpret polarized-light transitions between individual levels of the ion-crystal system.

\end{abstract}

\maketitle

\section{\label{sec:intro} Introduction}

The Judd-Ofelt (JO) theory has been successfully applied since almost 60 years, to interpret the intensities of absorption and emissions lines of crystals and glasses doped with trivalent lanthanide ions (Ln$^{3+}$) \cite{judd1962, ofelt1962, hehlen2013}. Despite its remarkable efficiency, this standard JO theory cannot reproduce some of the observed transitions, because of its strong selection rules. It is especially the case for europium (Eu$^{3+}$) \cite{tanner2013, binnemans2015}, well known to challenge the standard JO theory \cite{walsh2006}. Many extensions of the original model have been proposed to overcome this drawback \cite{smentek1998}, including \textit{e.g.}~J-mixing \cite{tanaka1994, kushida2002, kushida2003}, the Wybourne-Downer mechanism \cite{downer1988, burdick1989}, velocity-gauge expression of the electric-dipole (ED) operator \cite{smentek1997}, relativistic or configuration-interaction (CI) effects \cite{smentek2000, smentek2001, wybourne2002, ogasawara2005, dunina2008}, purely \textit{ab initio} intensity calculations \cite{wen2014}. In this respect, Smentek and coworkers were able to reproduce experimental absorption oscillator strengths with a very high accuracy, with up to 17 adjustable parameters \cite{kedziorski2007}. But in spite of all these improvements, even the most recent experimental studies use the standard version of the JO theory \cite{ciric2019a, ciric2019b}.

In the standard JO theory, the line strength characterizing a given transition is a linear combinations of three parameters $\Omega_\lambda$ (with $\lambda=2$, 4 and 6), which are functions of both the properties of the Ln$^{3+}$ ion and the crystal-field parameters \cite{walsh2006, hehlen2013}. Since the $\Omega_\lambda$-parameters are adjusted by least-square fitting, those two types of contributions cannot be separated. However, the properties of the impurity can be investigated by means of fee-ion spectroscopy. In this respect, recent joint experimental and theoretical investigations have provided a detailed knowledge of some free-Ln$^{3+}$ ion structure \cite{meftah2007, wyart2007, meftah2016, arab2019}. Although such a study has not been made with Eu$^{3+}$, the continuity of the atomic properties along the lanthanide series opens the possibility to compute the Eu$^{3+}$ spectrum using a semi-empirical method, based on adjusted parameters of neighboring elements \cite{wyart2011}.

In this article, we present a modified version of the JO theory in which the properties of the free Ln$^{3+}$ ions, \textit{i.e.}~energies and transition integrals, are computed using a combination of \textit{ab initio} and least-square fitting procedures available in Cowan's suite of codes \cite{cowan1981, kramida2019}. This allows us to relax some of the strong assumptions of the JO theory, for instance the strict application of the closure relation. The line strengths appear as linear combinations of three adjustable parameters which are only functions of the crystal-field potential, giving access to the local environment around the ion. We account for the spin-orbit (SO) interaction responsible of spin-changing transitions by calculating the line strengths at the third order of perturbation theory. Our results on Eu$^{3+}$ suggest that the spin-mixing transitions are mainly due to the SO mixing within the ground electronic configuration, in contradiction with the Wybourne-Downer mechanism described in Ref.~\cite{downer1988, burdick1989}. In addition, our model gives a simple physical interpretation of the transitions that are forbidden in the framework of the standard JO theory, including ${}^7F_0 \leftrightarrow {}^5D_0$, ${}^7F_0 \leftrightarrow {}^5D_J$ or ${}^7F_J \leftrightarrow {}^5D_0$ with $J$ odd. To benchmark our model, we reproduce quantitatively the set of experimental absorption oscillator strengths of Babu \textit{et al.}~\cite{babu2000}, although we overestimate the strength of ${}^7F_0 \leftrightarrow {}^5D_0$ transition.

The paper is organized as follows. Section \ref{sec:th} contains our analytical development resulting in the ED line strengths, which then allow for calculating oscillator strengths and Einstein coefficients. Our model is based on the time-independent perturbation theory, up to second and third orders (see Subsections \ref{sub:2nd} and \ref{sub:3rd} respectively). Then in Section \ref{sec:eu}, we apply our model to the case of europium, describing first the free-ion properties required for our model in Subsections \ref{sub:param}--\ref{sub:eu_transition}, and then the f-f transitions within the ground configuration in Subsection \ref{sub:ff_transitions}. Section \ref{sec:conclusion} contains conclusions and prospects.

\section{\label{sec:th} Electric-dipole line strengths}

The aim of the present section is to derive analytical expressions for the electric-dipole (ED) line strengths, which enable to characterize absorption and emission intensities of Ln$^{3+}$-doped solids. Unlike the magnetic-dipole (MD) and electric-quadrupole (EQ) transitions \cite{dodson2012}, the ED ones are activated by the presence of the host material, which relaxes the free-space selection rules. We use similar hypotheses as in the original JO model \cite{judd1962, ofelt1962}: the crystal-field (CF) potential slightly admixes the levels of the ground configuration [Xe]$4f^w$ and those of the first excited configuration [Xe]$4f^{w-1}5d$, where [Xe] denotes the ground configuration of xenon, dropped in the rest of the article. In the resulting perturbative expression of the ED line strength, we assume that all the levels of the excited configuration have the same energy. However, we relax some of the original hypothesis, by accounting for the energies of the ground-configuration levels, and by applying the closure relation less strictly. Unlike the standard and most common extensions of the JO model, we do not introduce effective operators, like the so-called unit-tensor operator $U^{(k)}$ \cite{cowan1981}, but rather work on the matrix elements of the CF or ED operators. To calculate the line strength, we firstly use the second-order perturbation theory (see subsection \ref{sub:2nd}) and then the third-order perturbation theory  (see subsection \ref{sub:3rd}), for which the free-ion spin-orbit operator is within the perturbation.

The common starting point of those two calculations is the multipolar expansion of the crystal-field potential,
\begin{equation}
  V_\mathrm{CF} = \sum_{kq} A_{kq} \sum_{j=1}^N r_j^k 
    C_q^{(k)}(\theta_j,\phi_j) \equiv \sum_{kq} A_{kq} P_q^{(k)}
  \label{eq:Vf}
\end{equation}
where $k$ is a non-negative integer and $q=-k$, $-k+1$, ..., $+k$, $(r_j,\theta_j,\phi_j)$ are the spherical coordinates of the $j$-th ($j=1$ to $N$) electron in the referential frame centered on the nucleus of the Ln$^{3+}$ ion, and $C_q^{(k)}$ are the Racah spherical harmonics of rank $k$ and component $q$, related to the usual spherical harmonics by $C_q^{(k)} (\theta_j,\phi_j) = \sqrt{4\pi/(2k+1)} \times Y_{kq} (\theta_j,\phi_j)$, see for example Chap.~5 of Ref.~\cite{varshalovich1988}. In Eq.~\eqref{eq:Vf}, the quantities $P_q^{(k)}$ represent the electric multipole moment as defined in Chaps.~14 and 15 of Ref.~\cite{cowan1981}. The simplest way of calculating the CF parameters $A_{kq}$ is to assume that they are due to distributed charges inside the host material. More elaborate models can be used, like distributed dipoles resulting in the so-called dynamical coupling \cite{smentek1998}, or the vibration of the ion center-of-mass. This would affect the physical origin of the $A_{kq}$ coefficients, but not the validity of the forthcoming results \cite{judd1962}.

\subsection{Second-order correction}
\label{sub:2nd}

In the theory of light-matter interaction, the ED approximation arises at the first order of perturbation theory. Furthermore, the f-f transitions in Ln$^{3+}$-doped solids are only possible if the free-ion levels are perturbed by the CF potential. Therefore, using the first-order correction on the ion levels to calculate the matrix element of the ED operator gives in total a second-order correction.

We call $|\Psi_i\rangle$ the eigenvectors associated with the ion+crystal system (without electromagnetic field). In the framework of perturbation theory, we express them as $|\Psi_i\rangle = \sum_m |\Psi_i^m\rangle$, where $m$ denotes the order of the perturbative expansion. In this subsection, we consider that the 0-th, \textit{i.e.}~unperturbed, eigenvectors $|\Psi_i^0\rangle$ are the free-ion levels. Those belonging to the ground configuration $n\ell^w$ (with $n\ell=4f$ for Ln$^{3+}$ ions) are written in intermediate coupling scheme \cite{cowan1981}
\begin{equation}
  |\Psi_i^0\rangle = \sum_{\alpha_i L_i S_i} c_{\alpha_i L_i S_i} 
   \left|n\ell^w\,\alpha_i L_i S_i J_i M_i \right\rangle
  \label{eq:0th-lev-gr}
\end{equation}
where $L_i$, $S_i$ and $J_i$ are the quantum numbers associated with the orbital, spin and total electronic angular momentum respectively, while $M_i$ is associated with the $z$-projection of the latter. The free-ion levels of energy $E_i^0 \equiv E_i$ are degenerate in $M_i$. Finally in Eq.~\eqref{eq:0th-lev-gr}, $\alpha$ is a generic notation containing additional information like the seniority number \cite{cowan1981}. In the $4f^w$ configuration of Ln$^{3+}$ ions, the energy levels are usually well described in the LS coupling scheme (see Table \ref{tab:Eu_ground}). 

In the first excited configuration $n\ell^{w-1}n'\ell'$, with $(n\ell,n'\ell') = (4f,5d)$ for Ln$^{3+}$ ions, we consider free-ion levels in pure LS coupling,
\begin{equation}
  |\Psi_t^0\rangle = |n\ell^{w-1} \overline{\alpha} \overline{L}
    \overline{S}, \, n'\ell' L'S'J'M' \rangle,
  \label{eq:0th-lev-exc}
\end{equation}
where the overlined quantum numbers characterize the $n\ell^{w-1}$ subshell alone. As Table \ref{tab:Eu_excited} shows, the LS coupling is not appropriate for the energy levels of the excited configuration. But since, in our ED matrix element calculation, we will assume that all the levels of the excited configuration have the same energy, the choice of coupling scheme is arbitrary, and so we take the simplest one.

Now we express the ED transition amplitude $D_{12}$ between eigenvectors $|\Psi_i^0\rangle + |\Psi_i^1\rangle$ ($i=1,2$), perturbed by the CF potential up to the first order,
\begin{align}
  D_{12} & = \left\langle \Psi_1^1 \right| P_p^{(1)}
    \left| \Psi_2^0 \right\rangle + \left\langle \Psi_1^0 \right| 
    P_p^{(1)} \left| \Psi_2^1 \right\rangle \nonumber \\
   & = \sum_{t} \left[ \frac{\left\langle \Psi_1^0 \right| 
     V_{\mathrm{CF}} \left| \Psi_t^0 \right \rangle } {E_1-E_t}
     \left\langle \Psi_t^0 \right|P_p^{(1)} \left|\Psi_2^0 
     \right\rangle \right. \nonumber \\
   & \qquad \left. + \left\langle \Psi_1^0 \right|P_p^{(1)}
     \left| \Psi_t^0 \right\rangle \frac{\left\langle \Psi_t^0 
     \right|V_{\mathrm{CF}} \left| \Psi_2^0\right\rangle } 
     {E_2-E_t} \right] \,,
  \label{eq:d12-1}
\end{align}
where the index $p=0$ denotes $\pi$ light polarization, and $p=\pm 1$ denote $\sigma^{\pm}$ polarizations. We recall that $\langle \Psi_1^0 |P_p^{(1)}| \Psi_2^0 \rangle = 0$, because in free space, there is no ED transition between levels of the same electronic configuration. In what follows, we assume that all the energies of the excited configuration are equal, $E_t \approx E_{n'\ell'}$. Rather than the center-of-gravity energy of the excited configuration, $E_t$ can be regarded as the mean energy for which the coupling with both levels 1 and 2 is significant (see Fig.~\ref{fig:log_gf}).

Equation \eqref{eq:d12-1} contains matrix elements of $P_p^{(1)}$ and $V_{\mathrm{CF}}$, which are themselves functions of $P_q^{(k)}$, as Eq.~\eqref{eq:Vf} shows. Being irreducible tensor operators, the matrix elements of $P_q^{(k)}$ satisfy the Wigner-Eckart theorem \cite{varshalovich1988}
\begin{equation}
  \left\langle \Psi_i^0 \right|P_q^{(k)} \left| \Psi_t^0 \right\rangle
   = \frac{C_{J'M'kq}^{J_iM_i}}{\sqrt{2J_i+1}} \left\langle \Psi_i^0 
     \right\Vert P^{(k)} \left\Vert \Psi_t^0 \right\rangle
  \label{eq:we}
\end{equation}
where $C_{J'M'kq}^{J_iM_i} = \langle J'M'kq | J'kJ_iM_i \rangle$ is a Clebsch-Gordan (CG) coefficient, and $\langle \Psi_i^0 \Vert P^{(k)} \Vert \Psi_t^0 \rangle$ the reduced matrix element given in Eq.~\eqref{eq:pq-red}, which is independent from $M_i$, $M'$ and $q$.

By contrast, the products of the kind $\langle \Psi_i^0 |P_q^{(k)} | \Psi_t^0 \rangle \langle \Psi_t^0 |P_p^{(1)} | \Psi_2^0 \rangle$ are not irreducible tensors; still we overcome this problem by expanding the product of two CG coefficients given in \cite{varshalovich1988}, which yields
\begin{align}
  & \left\langle \Psi_1^0 \right|P_q^{(k)} \left| \Psi_t^0 \right\rangle
    \left\langle \Psi_t^0 \right|P_p^{(1)} \left| \Psi_2^0 \right\rangle
  \nonumber \\
   & = \sum_{\lambda\mu} (-1)^{J_1+J_2-\lambda} \sqrt{ \frac{2\lambda+1} {2J_{1}+1}} C_{kq1p}^{\lambda\mu} C_{J_2M_2\lambda\mu}^{J_1M_1}
  \nonumber \\
   & \times \sixj{k}{1}{\lambda}{J_2}{J_1}{J'}
   \left\langle \Psi_1^0 \right\Vert P^{(k)} \left\Vert \Psi_t^0 \right\rangle \left\langle \Psi_t^0 \right\Vert P^{(1)} \left\Vert \Psi_2^0 \right\rangle
  \label{eq:prod-tens}
\end{align}
where the quantity between curly brackets is a Wigner 6-j symbol. Equation \eqref{eq:prod-tens} is interesting because the only dependence on quantum numbers $M_i$ is in the CG coefficient $C_{J_{2}M_{2}\lambda\mu} ^{J_{1}M_{1}}$, while $M'$ is absent. The equation appears as a sum of irreducible tensors of rank $\lambda$ and component $\mu$ coupling directly $|\Psi_1^0\rangle$ and $|\Psi_2^0\rangle$. The selection rules governing this coupling are $\Delta J=|J_{2}-J_{1}|\le\lambda\le J_{1}+J_{2}$ and $M_1=M_2+\mu$. Moreover, the triangle rule associated with $C_{kq1p}^{\lambda\mu}$ imposes $\lambda=k$, $k\pm 1$, $\mu=p+q$ and $-\lambda \le \mu \le +\lambda$.

Applying the same reasoning for the third line of Eq.~\eqref{eq:d12-1}, we obtain the same result as Eq.~(\ref{eq:prod-tens}) except the permutations of the couples of indexes $(k,q)$ and $(1,p)$. Using the symmetry relation of CG coefficients $C_{1pkq}^{\lambda\mu} = (-1)^{1+k-\lambda} C_{kq1p}^{\lambda\mu}$, we get to the final expression for the transition amplitude
\begin{widetext}
\begin{align}
  D_{12} & = \sum_{\alpha_{1}L_{1}S_{1}}c_{\alpha_{1}L_{1}S_{1}}
    \sum_{\alpha_{2}L_{2}S_{2}}c_{\alpha_{2}L_{2}S_{2}}
    \sum_{kq}\,A_{kq} \sum_{\lambda\mu} (-1)^{J_{1}+J_{2}-\lambda} 
    \sqrt{\frac{2\lambda+1}{2J_{1}+1}} C_{kq1p}^{\lambda\mu} 
    C_{J_{2}M_{2}\lambda\mu}^{J_{1}M_{1}} \nonumber \\
  & \times\sum_{J'} \left( \sixj{k}{1}{\lambda}{J_2}{J_1}{J'}
   \mathcal{D}_{12,J'}^{(k1)} + \left(-1\right)^{1+k-\lambda}
   \sixj{1}{k}{\lambda}{J_2}{J_1}{J'}
   \mathcal{D}_{12,J'}^{(1k)}\right),
  \label{eq:d12-2}
\end{align}
where we have introduced the quantities
\begin{align}
  \mathcal{D}_{12,J'}^{(k1)} & = \frac{1}{E_1-E_{n'\ell'}}
    \sum_{\overline{\alpha} \overline{L} \overline{S},\, L'S'J'}
    \left\langle n\ell^w\,\alpha_1L_1S_1J_1 \right\Vert P^{(k)} \left\Vert \overline{\alpha} \overline{L} \overline{S},\, L'S'J' \right\rangle \left\langle \overline{\alpha} \overline{L} \overline{S}, \, L'S'J' \right\Vert P^{(1)} \left\Vert  n\ell^w\,\alpha_2L_2S_2J_2 \right\rangle
  \label{eq:dk1} \\
  \mathcal{D}_{12,J'}^{(1k)} & = \frac{1}{E_2-E_{n'\ell'}}
    \sum_{\overline{\alpha} \overline{L} \overline{S},\, L'S'J'}
    \left\langle n\ell^w\,\alpha_1L_1S_1J_1 \right\Vert P^{(1)} \left\Vert \overline{\alpha} \overline{L} \overline{S},\, L'S'J' \right\rangle \left\langle \overline{\alpha} \overline{L} \overline{S},\, L'S'J' \right\Vert P^{(k)} \left\Vert  n\ell^w\,\alpha_2L_2S_2J_2 \right\rangle
  \label{eq:d1k}
\end{align}
\end{widetext}
in which $| \overline{\alpha} \overline{L} \overline{S} ,\, L'S'J' \rangle$ is a condensed representation of $|\Psi_t^0 \rangle$, see Eq.~\eqref{eq:0th-lev-exc}. The superscripts $(k1)$ and $(1k)$ correspond to the order in which the tensor operators $P^{(k)}$ and $P^{(1)}$ are written. 

For eigenvectors $|\Psi_{1,2}^0 \rangle$ belonging to the ground configuration and $|\Psi_t^0 \rangle$ belonging to the first excited configuration, the 3-j symbol of Eq.~\eqref{eq:pq-red} imposes that the CF potential matrix elements are non-zero for $k=1$, 3 and 5, which, according to Eq.~\eqref{eq:prod-tens}, imposes $\lambda=0$, 1, ..., 6. By contrast, in the standard version of the JO theory, $\lambda = k+1 = 2$, 4 and 6. The $\lambda=0$ contribution in Eq.~\eqref{eq:prod-tens} comes from the dipolar term $k=1$ of the CF potential; it is the only non-zero contribution when $J_1=J_2=0$, for instance the $^5D_0 \leftrightarrow {}^7F_0$ transition in Eu$^{3+}$. Our odd-$\lambda$ contributions are responsible for the transitions like $^5D_0 \leftrightarrow {}^7F_{3,5}$ and $^5D_3 \leftrightarrow {}^7F_0$; they arise because we consider distinct energies for levels 1 and 2, $E_1 \neq E_2$, unlike the standard JO theory. But since the energy difference $|E_2 - E_1|$ is significantly smaller (although not negligible) compared to $E_{n'\ell'} - E_{1,2}$, those transitions are weak. Finally, since the operators $P^{(k)}$ do not couple different spin states, the spin-changing transitions are only due to the mixing of different spin states within the ground-configuration levels $|\Psi_{1,2}^0\rangle$, see Eq.~\eqref{eq:0th-lev-gr}. In other words, Eq.~\eqref{eq:d12-2} does not account for the so-called Downer-Wybourne mechanism \cite{downer1988}.

At present, we calculate the ED line strength $\mathcal{S}_\mathrm{ED} = \sum_{pM_1M_2} (D_{12})^2$. Expressing Eq.~\eqref{eq:d12-2} twice gives many sums: in particular on $p$, $M_1$, $M_2$, $k$, $q$, $\mu$ and $J'$, but also $k'$, $q'$, $\mu'$ and $J''$ (coming from the second expansion of $D_{12}$). Focusing on the sum involving CG coefficients, we have
\begin{align}
  & \sum_{pqq'\mu\mu'} C_{kq1p}^{\lambda\mu} C_{k'q'1p}^{\lambda'\mu'} \sum_{M_{1}M_{2}} \frac{ C_{J_{2}M_{2}\lambda\mu}^{J_{1}M_{1}} C_{J_{2}M_{2}\lambda'\mu'}^{J_{1}M_{1}}} {2J_{1}+1}
  \nonumber \\
  & = \sum_{pqq'\mu\mu'} C_{kq1p}^{\lambda\mu} C_{k'q'1p}^{\lambda'\mu'} \frac{\delta_{\lambda\lambda'}\delta_{\mu\mu'}} {2\lambda+1}
  \nonumber \\
  & = \frac{\delta_{\lambda\lambda'}}{2\lambda+1} \sum_{pqq'\mu}  C_{kq1p}^{\lambda\mu} C_{k'q'1p}^{\lambda\mu} =  \frac{\delta_{\lambda\lambda'} \delta_{kk'} \delta_{qq'}} {2k+1} \,, 
  \label{eq:sum-cg}
\end{align}
where the Kronecker symbols come from the orthonormalization relation of CG coefficients. Plugging Eq.~\eqref{eq:sum-cg} into the line strength gives
\begin{widetext}
\begin{align}
  \mathcal{S}_{\mathrm{ED}} & = \sum_{\alpha_{1a}L_{1a}S_{1a}} 
   c_{\alpha_{1a}L_{1a}S_{1a}} \sum_{\alpha_{2a}L_{2a}S_{2a}} c_{\alpha_{2a}L_{2a}S_{2a}} \sum_{\alpha_{1b}L_{1b}S_{1b}} c_{\alpha_{1b}L_{1b}S_{1b}} \sum_{\alpha_{2b}L_{2b}S_{2b}} c_{\alpha_{2b}L_{2b}S_{2b}} \sum_{kq} \frac{|A_{kq}|^{2}}{2k+1} 
  \nonumber \\
  & \times \sum_{\lambda} \left(2\lambda+1\right) \sum_{J'}
   \left( \sixj{k}{1}{\lambda}{J_2}{J_1}{J'}
   \mathcal{D}_{1a,2a,J'}^{(k1)} + \left(-1\right)^{1+k-\lambda}
    \sixj{1}{k}{\lambda}{J_2}{J_1}{J'} \mathcal{D}_{1a,2a,J'}^{(1k)} 
  \right) \nonumber \\
  & \times \sum_{J''} \left( \sixj{k}{1}{\lambda}{J_2}{J_1}{J''} 
   \mathcal{D}_{1b,2b,J''}^{(k1)} +\left(-1\right)^{1+k-\lambda}
   \sixj{1}{k}{\lambda}{J_2}{J_1}{J''} \mathcal{D}_{1b,2b,J''}^{(1k)} \right) .
\end{align}
When expanded, the last two lines contain four terms: two of the kind
\begin{equation}
  \sum_{\lambda} \left(2\lambda+1\right)
   \sixj{k_1}{k_2}{\lambda}{J_2}{J_1}{J'}
   \sixj{k_1}{k_2}{\lambda}{J_2}{J_1}{J''}
   \mathcal{D}_{1a,2a,J'}^{(k_1k_2)} \mathcal{D}_{1b,2b,J''}^{(k_1k_2)} 
  = \frac{\delta_{J'J''}} {2J'+1} \mathcal{D}_{1a,2a,J'}^{(k_1k_2)} 
   \mathcal{D}_{1b,2b,J'}^{(k_1k_2)}
  \label{eq:pr-6j-1}
\end{equation}
where the sum on $\lambda$ is actually the orthonormalization relation of
6-j symbols; and two terms of the kind
\begin{align}
  & \sum_{\lambda} \left(-1\right)^{k_{1}+k_{2}-\lambda}
   \left(2\lambda+1\right)
   \sixj{k_1}{k_2}{\lambda}{J_2}{J_1}{J'}
   \sixj{k_2}{k_1}{\lambda}{J_2}{J_1}{J''}
   \mathcal{D}_{1a,2a,J'}^{(k_{1}k_{2})}
   \mathcal{D}_{1b,2b,J''}^{(k_{2}k_{1})}
  \nonumber \\
  & \quad = \left(-1\right)^{k_{1}+k_{2}+J'+J''} 
   \sixj{k_1}{J_1}{J'}{k_2}{J_2}{J''}
   \mathcal{D}_{1a,2a,J'}^{(k_{1}k_{2})}
   \mathcal{D}_{1b,2b,J''}^{(k_{2}k_{1})}
  \label{eq:pr-6j-2}
\end{align}
where we use some properties of 6-j symbols (see Ref.~\cite{varshalovich1988}, p.~305). The final expression of the line strength is
\begin{align}
  \mathcal{S}_{\mathrm{ED}} & = \sum_{\alpha_{1a}L_{1a}S_{1a}} 
   c_{\alpha_{1a}L_{1a}S_{1a}} \sum_{\alpha_{2a}L_{2a}S_{2a}} c_{\alpha_{2a}L_{2a}S_{2a}} \sum_{\alpha_{1b}L_{1b}S_{1b}} c_{\alpha_{1b}L_{1b}S_{1b}} \sum_{\alpha_{2b}L_{2b}S_{2b}} c_{\alpha_{2b}L_{2b}S_{2b}} \sum_{kq}\frac{|A_{kq}|^{2}}{2k+1} 
  \nonumber \\
  & \times \sum_{J'} \left[ \frac{ \mathcal{D}_{1a,2a,J'}^{(k1)} \mathcal{D}_{1b,2b,J'}^{(k1)} + \mathcal{D}_{1a,2a,J'}^{(1k)} \mathcal{D}_{1b,2b,J'}^{(1k)}}{2J'+1}
   + \sum_{J''} \left(-1\right)^{1+k+J'+J''} \right. \nonumber \\
  & \left. \phantom{\sum_{J'}\frac{\mathcal{D}_{,J'}^{(}}{J'}}
   \times\left( \sixj{k}{J_1}{J'}{1}{J_2}{J''}
   \mathcal{D}_{1a,2a,J'}^{(k1)} \mathcal{D}_{1b,2b,J''}^{(1k)}
   + \sixj{1}{J_1}{J'}{k}{J_2}{J''} \mathcal{D}_{1a,2a,J'}^{(1k)} 
   \mathcal{D}_{1b,2b,J''}^{(k1)} \right) \right]
  \label{eq:sed-ext-1}
\end{align}
\end{widetext}

Equation (\ref{eq:sed-ext-1}) looks very different from the standard JO line strength $\mathcal{S}_\mathrm{ED} = \sum_\lambda \Omega_\lambda \langle \Psi_1^0 \Vert U^{(\lambda)} \Vert \Psi_2^0 \rangle$, especially because it does not depend on $\lambda$, but depends on $J'$ and $J''$ (which are by contrast eliminated in the standard case). The index $\lambda$ is still relevant in the ED transition amplitude $D_{12}$, see Eq.~\eqref{eq:d12-2}, because it allows for deriving the selection rules, but it disappears in the line strength, where we consider unpolarized light and ions (that is to say sums on $p$, $M_{1}$ and $M_{2}$). In Eq.~(\ref{eq:sed-ext-1}), the influence of the CF potential are only contained in the three parameters $X_{k} = (2k+1)^{-1} \sum_{q} |A_{kq}|^{2}$, for $k=1$, 3 and 5, which are $q$-averages of the square of $A_{kq}$. In what follows, they will be treated as adjustable parameters, whereas all the atomic properties will be computed using atomic-structure methods.

\subsection{Third-order correction}
\label{sub:3rd}

In this section, we address the influence of spin-orbit (SO) mixing in the excited configuration on spin-changing f-f transitions. Contrary to the ground configuration, the LS coupling scheme is by far not appropriate to interpret the levels of the $4f^{w-1}5d$ configuration (see Table \ref{tab:Eu_excited}), because the electrostatic energy between $4f$ and $5d$ electrons and the SO energy of the $5d$ electron are comparable. Therefore one can expect these excited levels to play a significant role in the spin-changing transitions.

To check this hypothesis, we will investigate the effect of the SO Hamiltonian of the ion $H_\mathrm{SO}$ using perturbation theory. Namely we define a perturbation operator $V$ containing SO and CF interactions,
\begin{equation}
  V = H_\mathrm{SO} + V_\mathrm{CF}\,.
\end{equation}
In consequence, the new unperturbed eigenvectors related to the ground configuration are called manifolds, \textit{i.e.}~atomic levels for which the SO energy is set to 0. Those manifolds $|\widetilde{\Psi}_i^0 \rangle$, of energy $\widetilde{E}_i$ are degenerate in $M_i$ as previously, but also in $J_i$, and they are characterized by one $L_i$ and one $S_i$ quantum number,
\begin{equation}
  |\widetilde{\Psi}_i^0 \rangle = \sum_{\alpha_i} \widetilde{c}_{\alpha_i} 
   \left|n\ell^w\,\alpha_i L_i S_i J_i M_i \right\rangle
  \label{eq:0th-man-gr}
\end{equation}
Some manifolds, like the lowest $^5$D one in Eu$^{3+}$, are linear combination of different terms having the same $L$ and $S$ but different seniority numbers, hence the sum on $\alpha$ in Eq.~\eqref{eq:0th-man-gr}. For the excited configuration, the unperturbed eigenvectors are those given in Eq.~\eqref{eq:0th-lev-exc}.

The selection rules associated with $H_\mathrm{SO}$ and $V_\mathrm{CF}$ are very different. In particular, $H_\mathrm{SO}$ couples unperturbed eigenvectors of the same configuration, whereas the odd terms of $V_\mathrm{CF}$ couple configurations of opposite parities. Therefore, the influence of both SO and CF potentials appears as products of matrix elements like $\langle \widetilde{\Psi}_1^0 |H_\mathrm{SO} |\widetilde{\Psi}_i^0 \rangle \langle \widetilde{\Psi}_i^0 |V_\mathrm{CF} |\Psi_t^0 \rangle$, and we need to go to the third order of perturbation theory to calculate the transition amplitude,
\begin{equation}
  D_{12} = \left\langle \widetilde{\Psi}_1^1 \right| P_p^{(1)}
   \left| \widetilde{\Psi}_2^1 \right\rangle + \left\langle
   \widetilde{\Psi}_1^2 \right| P_p^{(1)} \left|
   \widetilde{\Psi}_2^0 \right\rangle + \left\langle \widetilde{\Psi}_1^0 
   \right| P_p^{(1)} \left| \widetilde{\Psi}_2^2 \right\rangle
  \label{eq:d12-3}
\end{equation}
where the second-order correction of eigenvectors is given in Eq.~\eqref{eq:2nd-eigvct}.

By expanding Eq.~\eqref{eq:d12-3}, we get six terms corresponding to the six possible products of matrix element of $H_\mathrm{SO}$, $P_p^{(1)}$ and $P_q^{(k)}$. Since $H_\mathrm{SO}$ couples states of the same configuration, unlike $P_p^{(1)}$ and $P_q^{(k)}$, we distinguish two kinds of terms:
\begin{itemize}
  \item $\langle \widetilde{\Psi}_1^0 |P_q^{(k)} |\Psi_t^0 \rangle \times \langle \Psi_t^0  | H_\mathrm{SO} | \Psi_u^0 \rangle \times \langle \Psi_u^0 | P_p^{(1)} | \widetilde{\Psi}_2^0 \rangle$ and $\langle \widetilde{\Psi}_1^0 |P_p^{(1)} |\Psi_t^0 \rangle \times \langle \Psi_t^0  | H_\mathrm{SO} | \Psi_u^0 \rangle \times \langle \Psi_u^0 | P_q^{(k)} | \widetilde{\Psi}_2^0 \rangle$, for which the SO interaction mixes levels of the excited configuration, for example quintet and septets in Eu$^{3+}$;
  \item $\langle \widetilde{\Psi}_1^0 | H_\mathrm{SO} | \widetilde{\Psi}_{i'_1}^0 \rangle \times \langle \widetilde{\Psi}_{i'_1}^0 | P_q^{(k)} | \Psi_t^0 \rangle \times \langle \Psi_t^0 | P_p^{(1)} | \widetilde{\Psi}_2^0 \rangle$, $\langle \widetilde{\Psi}_1^0 | H_\mathrm{SO} | \widetilde{\Psi}_{i'_1}^0 \rangle \times \langle \widetilde{\Psi}_{i'_1}^0 | P_p^{(1)} | \Psi_t^0 \rangle \times \langle \Psi_t^0 | P_q^{(k)} | \widetilde{\Psi}_2^0 \rangle$, 
  $\langle \widetilde{\Psi}_1^0 | P_q^{(k)} | \Psi_t^0 \rangle \times \langle \Psi_t^0 | P_p^{(1)} | \widetilde{\Psi}_{i'_2}^0 \rangle \times \langle \widetilde{\Psi}_{i'_2}^0 | H_\mathrm{SO} | \widetilde{\Psi}_2^0 \rangle$ and $\langle \widetilde{\Psi}_1^0 | P_p^{(1)} | \Psi_t^0 \rangle \times \langle \Psi_t^0 | P_q^{(k)} | \widetilde{\Psi}_{i'_2}^0 \rangle \times \langle \widetilde{\Psi}_{i'_2}^0 | H_\mathrm{SO} | \widetilde{\Psi}_2^0 \rangle$. In those cases, the SO interaction mixes manifolds of the ground configuration, for example in Eu$^{3+}$, $^7$F with $^5$D, $^5$F and $^5$G.
\end{itemize}

Because $H_\mathrm{SO}$ is a scalar, \textit{i.e.}~a tensor operator of rank 0, the application of the Wigner-Eckart theorem gives a CG coefficients $C_{JM00}^{J'M'} =\delta_{JJ'} \delta_{MM'}$. So, applying the Wigner-Eckart theorem to $P_q^{(k)}$ and $P_p^{(1)}$ as in Eq.~\eqref{eq:we}, the product of three matrix elements can be expanded in a similar way to Eq~\eqref{eq:prod-tens}. For example,
\begin{align}
  & \left\langle \Psi_1^0 \right| P_q^{(k)} \left| \Psi_t^0 \right\rangle
    \left\langle \Psi_t^0 \right| H_\mathrm{SO} \left| \Psi_u^0 \right\rangle
    \left\langle \Psi_u^0 \right| P_p^{(1)} \left| \Psi_2^0 \right\rangle
  \nonumber \\
   & = \sum_{\lambda\mu} (-1)^{J_1+J_2-\lambda} \sqrt{ \frac{2\lambda+1} {2J_{1}+1}} C_{kq1p}^{\lambda\mu} C_{J_2M_2\lambda\mu}^{J_1M_1}
  \nonumber \\
   & \times \sixj{k}{1}{\lambda}{J_2}{J_1}{J'}
   \left\langle \Psi_1^0 \right\Vert P^{(k)} \left\Vert \Psi_t^0 \right\rangle
   \left\langle \Psi_t^0 \right| H_\mathrm{SO} \left| \Psi_u^0 \right\rangle
   \left\langle \Psi_u^0 \right\Vert P^{(1)} \left\Vert \Psi_2^0 \right\rangle
  \label{eq:prod-tens-2}
\end{align}
where $| \Psi_{t,u}^0 \rangle$ are two eigenvectors of the excited configuration with the same total angular momentum $J'$. The other products give similar results: the order of reduced matrix elements in the last line is of course the same as the order of matrix elements in the first line; if $P_p^{(1)}$ appears before $P_q^{(k)}$, the $1$ and $k$ are interchanged in the CG coefficients and 6-j symbols, like in Eq.~(\ref{eq:d12-2}).

Gathering the six matrix-element products, we can write the ED transition amplitude as
\begin{widetext}
\begin{align}
  D_{12} & = \sum_{\alpha_1 \alpha_2} \widetilde{c}_{\alpha_1}  \widetilde{c}_{\alpha_2} \sum_{kq}\, A_{kq} \sum_{\lambda\mu} (-1)^{J_{1}+J_{2}-\lambda} \sqrt{\frac{2\lambda+1}{2J_1+1}} C_{kq1p}^{\lambda\mu} C_{J_{2}M_{2}\lambda\mu}^{J_1M_1}
  \nonumber \\
  & \times \sum_{J'} \left[ \sixj{k}{1}{\lambda}{J_2}{J_1}{J'}
   \left(\mathcal{D}_{12,J'}^{(0k1)} + \mathcal{D}_{12,J'}^{(k01)} + \mathcal{D}_{12,J'}^{(k10)} \right) \right.
  \nonumber \\
  & \quad \left. + \left(-1\right)^{1+k-\lambda} 
   \sixj{1}{k}{\lambda}{J_2}{J_1}{J'} \left( \mathcal{D}_{12,J'}^{(01k)} + \mathcal{D}_{12,J'}^{(10k)} + \mathcal{D}_{12,J'}^{(1k0)} \right) \right] , 
  \label{eq:d12-4}
\end{align}
\end{widetext}
where the terms $\mathcal{D}_{12,J'}^{(k_{1}k_{2}k_{3})}$ are built in analogy to Eqs.~(\ref{eq:dk1}) and (\ref{eq:d1k}): the order of the superscripts is the one in which the matrix element of operators appear (the ``{}0" standing for $H_\mathrm{SO}$). Firstly, the $\mathcal{D}_{12,J'}^{(0k_{2}k_{3})}$ are such that
\begin{align}
  \mathcal{D}_{12,J'}^{(0k_{2}k_{3})} & 
   = \frac{1}{\Delta_{k_{2}k_{3}}} \sum_{i'_1 \alpha'_1 \beta'_1 L'_1}
    \frac{ \widetilde{c}_{\alpha'_1} \widetilde{c}_{\beta'_1} }
    {\widetilde{E}_1-\widetilde{E}_{i'_1}} 
   \nonumber \\
  & \times \left\langle \alpha_1 L_1 S_1 J_1 \right| H_{\mathrm{SO}} \left| \alpha'_1 L'_1 S_2 J_1 \right\rangle 
  \nonumber \\
  & \times \sum_{\overline{\alpha} \overline{L} \overline{S},L'_{2}} 
   \left\langle \beta'_1 L'_1 S_2 J_1 \right\Vert P^{(k_{2})} \left\Vert  \overline{\alpha} \overline{L} \overline{S}, L'_{2}S_{2}J' \right\rangle
  \nonumber \\
  & \times \left\langle \overline{\alpha} \overline{L} \overline{S}, L'_{2}S_{2}J' \right\Vert P^{(k_{3})} \left\Vert \alpha_2 L_{2}S_{2}J_{2} \right\rangle ,
  \label{eq:D1}
\end{align}
with
\begin{equation}
  \Delta_{lm} = \begin{cases}
   \widetilde{E}_{1}-E_{n'\ell'} & \mathrm{for\,\,}(l,m)=(k,1)\\
   \widetilde{E}_{2}-E_{n'\ell'} & \mathrm{for\,\,}(l,m)=(1,k).
\end{cases}\label{eq:d-dlt-k1k2}
\end{equation}
In Eu$^{3+}$ for example, for $|\widetilde{\Psi}_1^0 \rangle$ in the lowest $^{5}$D manifold and $|\widetilde{\Psi}_1^0 \rangle$ in the $^{7}$F manifold, $H_{\mathrm{SO}}$ couples $|\widetilde{\Psi}_1^0 \rangle$ to the $^{7}L'_{1}$ manifolds on the ground configuration (actually there is only one: $^{7}$F). The quantum numbers $(\overline{\alpha}\overline{L}\overline{S},L'_{2})$ characterize the septet levels ($S_2=3$) of the excited configuration. Similarly,
\begin{align}
  \mathcal{D}_{12,J'}^{(k_{1}k_{2}0)} & = \frac{1}{\Delta_{k_{1}k_{2}}} 
   \sum_{\overline{\alpha}\overline{L}\overline{S},L'_{1}}
   \left\langle \alpha_1 L_1 S_1 J_1 \right\Vert P^{(k_1)} \left\Vert \overline{\alpha} \overline{L} \overline{S},L'_{1}S_{1}J' \right\rangle
  \nonumber \\
   & \times \sum_{i'_2 \alpha'_2 \beta'_2 L'_2} \frac{\widetilde{c}_{\alpha'_2} \widetilde{c}_{\beta'_2}}{\widetilde{E}_2-\widetilde{E}_{i'_2}}
  \nonumber \\
  & \times \left\langle \overline{\alpha} \overline{L} \overline{S}, L'_{1}S_{1}J' \right\Vert P^{(k_2)} \left\Vert \alpha'_2 L'_2 S_1 J_2 \right\rangle
  \nonumber \\
  & \times \left\langle \beta'_2 L'_2 S_1 J_2 \right| H_{\mathrm{SO}} \left|\alpha_2 L_2 S_2 J_2 \right\rangle ,
  \label{eq:D1b}
\end{align}
where $\Delta_{k_{1}k_{2}}$ is given by Eq.~(\ref{eq:d-dlt-k1k2}). Here the SO Hamiltonian couples the $^{7}F$ manifold to the various quintet manifolds, for instance $^{5}D$, $^{5}F$ and $^{5}G$ manifolds, since $L'_2-L_2 = 0, \, \pm 1$. Finally, the terms $\mathcal{D}_{12,J'}^{(k_{1}0k_{3})}$ correspond to the Wybourne-Downer mechanism \cite{downer1988} where $H_{\mathrm{SO}}$ couples the quintet and septet levels of the excited configuration. Namely,
\begin{align}
  \mathcal{D}_{12,J'}^{(k_{1}0k_{3})} & = \frac{1}{\Delta_{k_{1}k_{3}}^{2}} 
   \sum_{\overline{\alpha} \overline{L} \overline{S}} \sum_{L'_{1}L'_{2}} \left\langle \alpha_{1}L_{1}S_{1}J_{1} \right\Vert P^{(k_{1})} \left\Vert \overline{\alpha} \overline{L} \overline{S}, L'_{1}S_{1}J' \right\rangle 
  \nonumber \\
  & \times \left\langle \overline{\alpha} \overline{L} \overline{S}, L'_{1}S_{1}J' \right| H_{\mathrm{SO}} \left| \overline{\alpha} \overline{L} \overline{S}, L'_{2}S_{2}J' \right\rangle
  \nonumber \\
  & \times \left\langle \overline{\alpha} \overline{L} \overline{S}, L'_{2}S_{2}J' \right\Vert P^{(k_{3})} \left\Vert \alpha_2 L_{2}S_{2}J_{2} \right\rangle ,
  \label{eq:D2}
\end{align}
where $\Delta_{k_{1}k_{3}}$ is given by Eq.~(\ref{eq:d-dlt-k1k2}).

If we assume that in Equations (\ref{eq:D1}), (\ref{eq:D1b}) and (\ref{eq:D2}), the spin-orbit interactions are of the same order of magnitude (see Table \ref{tab:param_sf}), the main difference between them comes from the energy denominator. The quantity $\Delta_{lm}$ is on the order of several tens of thousands of cm$^{-1}$, while the differences $\widetilde{E}_1-\widetilde{E}_{i'_1}$ and $\widetilde{E}_2-\widetilde{E}_{i'_2}$, which are the energies between different manifolds of the ground configuration, are on the order of several thousand cm$^{-1}$. This means that Eq.~(\ref{eq:D2}) is, roughly speaking, one order of magnitude smaller than Eqs.~(\ref{eq:D1}) and (\ref{eq:D1b}). This fact is really a precious information that brings the third order correction.

Combining Eqs.~(\ref{eq:sed-ext-1}) and (\ref{eq:d12-4}), we can see that the ED line strength $\mathcal{S}_{\mathrm{ED}}$ now contains 36 terms, containing products of the kind $\mathcal{D}_{12,J'}^{(k_{1a}k_{2a}k_{3a})} \times \mathcal{D}_{12,J''}^{(k_{1b}k_{2b}k_{3b})}$. For the 18 terms in which $k$ and 1 appear in the same order in $(k_{1a}k_{2a}k_{3a})$ and $(k_{1b}k_{2b}k_{3b})$, we have the same prefactor as the second line of Eq.~(\ref{eq:sed-ext-1}), that is $(2J'+1)^{-1}$. For the 18 other terms in which $k$ and 1 appear in different orders, we have the prefactors with the 6-j symbols as in the second and third lines of Eq.~(\ref{eq:sed-ext-1}). Namely, we can write the line strength as
\begin{widetext}
\begin{align}
  \mathcal{S}_{\mathrm{ED}} & = \sum_{\alpha_{1a} \alpha_{1b}} \widetilde{c}_{\alpha_{1a}} \widetilde{c}_{\alpha_{1b}} \sum_{\alpha_{2a} \alpha_{2b}} \widetilde{c}_{\alpha_{2a}} \widetilde{c}_{\alpha_{2b}} \sum_{kq} \frac{|A_{kq}|^{2}}{2k+1} \sum_{\kappa_{a}\kappa_{b}} \sum_{J'} \left[\frac{\widetilde{\delta}_{\kappa_{a},(k1)}\widetilde{\delta}_{\kappa_{b},(k1)}+\widetilde{\delta}_{\kappa_{a},(1k)}\widetilde{\delta}_{\kappa_{b},(1k)}}{2J'+1}\,\mathcal{D}_{1a,2a,J'}^{(\kappa_{a})}\mathcal{D}_{1b,2b,J'}^{(\kappa_{b})}\right.\nonumber \\
 & \left.\quad+\sum_{J''}\left(-1\right)^{1+k+J'+J''}\left(\left\{ \begin{array}{ccc}
k & J_{1} & J'\\
1 & J_{2} & J''
\end{array}\right\} \widetilde{\delta}_{\kappa_{a},(k1)}\widetilde{\delta}_{\kappa_{b},(1k)}+\left\{ \begin{array}{ccc}
1 & J_{1} & J'\\
k & J_{2} & J''
\end{array}\right\} \widetilde{\delta}_{\kappa_{a},(1k)}\widetilde{\delta}_{\kappa_{b},(k1)}\right)\mathcal{D}_{1a,2a,J'}^{(\kappa_{a})}\mathcal{D}_{1b,2b,J''}^{(\kappa_{b})}\right]\label{eq:sed-ext-2}
\end{align}
\end{widetext}
where $\kappa_{a} = (k_{1a} k_{2a} k_{3a})$ and $\kappa_{b} = (k_{1b} k_{2b} k_{3b})$ designate the possible combinations of indices $k$, 1 and 0. The quantity $\widetilde{\delta}_{\kappa,(k1)}=1$ if $\kappa$ is a combination in which $k$ appears firstly and 1 secondly, namely $\kappa=(k10)$, $(k01)$, $(0k1)$, and 0 otherwise. The quantity $\widetilde{\delta}_{\kappa,(1k)}$ corresponds to the inverse situation. Similarly to Eq.~\eqref{eq:sed-ext-1}, the line strength \eqref{eq:sed-ext-1} depends on the CF potential through the three parameters $X_k = (2k+1)^{-1} \sum_{q} |A_{kq}|^{2}$, which will be treated as adjustable in the next section.

\section{\label{sec:eu}Application to Europium}

In this section, we aim at benchmarking our model with experimental data. To this end, we have chosen the measurements of absorption oscillator strengths by Babu \textit{et al.}~\cite{babu2000}, who performed a thorough spectroscopic study of Eu$^{3+}$-doped lithium borate and lithium fluoroborate glasses. Because our model relies on free-ion properties, we start with studying the free-ion energies of the two lowest Eu$^{3+}$ electronic configurations $4f^6$, of even parity, and $4f^5 5d$, of odd parity, and the free-space transitions between them.

\subsection{\label{sub:param}Calculation of free-ion energy parameters}

The calculations of the Eu$^{3+}$ free-ion spectrum are performed with the semi-empirical technique provided by Robert Cowan's atomic-structure suite of codes \cite{kramida2019}, whose theoretical background is presented in Ref.~\cite{cowan1981}. In a first step, \textit{ab initio} radial wave functions $P_{n\ell}$ for all the subshells $n\ell$ of the considered configurations are computed with the relativistic Hartree-Fock (HFR) method. Those wave functions are used to calculate energy parameters, for instance center-of-gravity configuration energies $E_\mathrm{av}$, direct $F^k$ and exchange $G^k$ electrostatic integrals, or spin-orbit integrals $\zeta_{n\ell}$, that are the building blocks of the atomic Hamiltonian. In a second step, the latter are treated as adjustable parameters of a least-sqaure fitting calculation, in order to find the best possible agreement between the Hamiltonian eigenvalues and the experimental energies. To make some comparisons between different elements and ionization stages, one often defines the scaling factor (SF) $f_X = X_\mathrm{fit} / X_\mathrm{HFR}$ between the fitted and the HFR value of a given parameter $X$.

In an attempt to improve the quality of the fit (and therefore, the accuracy of the resulting eigenvectors), a variety of ``effective-operator" parameters, called $\alpha$, $\beta$ and $\gamma$ and {}``illegal''-k $F^k$, $G^k$ have been introduced, representing corrections to both the electrostatic and the magnetic single-configuration effects \cite{cowan1981}. {}``Illegal''-k means that these are the values of $k$ for which $k + \ell + \ell'$ is odd. These effective parameters, unlike other parameters, can not be calculated \textit{ab initio}. By contrast, we do not include the $M^k$, $T^k$ and $P^k$ parameters that are sometimes used in Ln$^{3+}$-ion ground configuration.
The general methodology for our fitting calculations is as follows: (a) fitting the parameters with an \textit{ab initio} values while effective parameters are forced to be zero; (b) fixing the parameters resulting from step (a) and fitting the effective parameters; (c) using the final values of (b), fitting all the parameters together.

\begin{figure}
\includegraphics[scale=0.63]{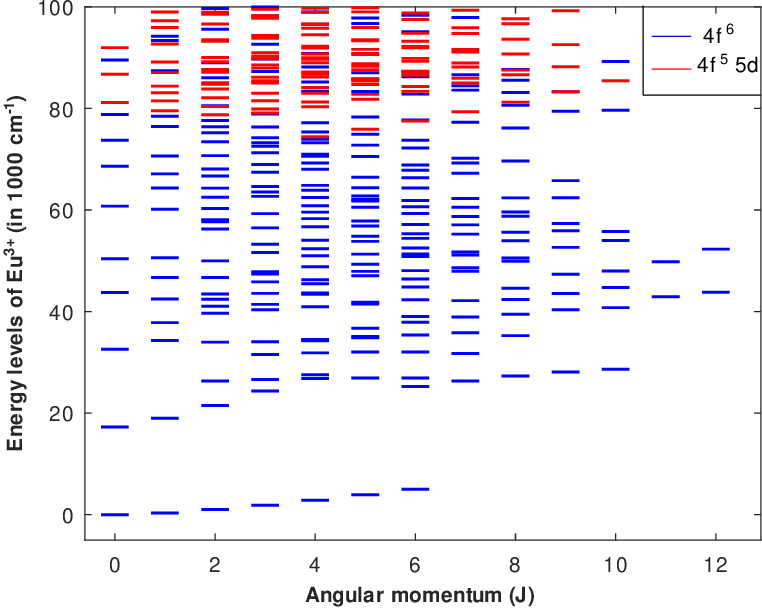}
\caption{\label{fig:energy_scheme} Energy levels of the $4f^{6}$ (blue) and $4f^{5} 5d$ (red) configurations of Eu$^{3+}$ as functions of the electronic angular momentum $J$. The plot is restricted to energy values between 0 and 100000~cm$^{−1}$.}
\end{figure}

Our fitting calculations require experimental energies. For the Eu$^{3+}$ ground configuration $4f^6$, we find them on the NIST ASD database \cite{NIST_ASD}. However, no experimental level has been reported for the $4f^5 5d$ configuration. Because the $4f^w$ configurations (with $2 \le w \le 12$) and the $4f^{w-1} 5d$ ones (with $3 \le w \le 13$) possess the same energy parameters, we perform a least-square fitting calculation of some $4f^{w-1}5d$ configurations for which experimental levels are known, namely for Nd$^{3+}$ ($w=3$) and Er$^{3+}$ ($w=11$) \cite{wyart2007, meftah2016, arab2019}.  Then, relying on the regularities of the scaling factors $f_X$ along the lanthanide series, we multiply the obtained scaling factors given in Table \ref{tab:param_sf} by the HFR parameters for Eu$^{3+}$ to compute the energies of $4f^5 5d$.

\begin{table*}
\caption{\label{tab:param_sf} Scaling factors (SFs) of \textit{ab initio} parameters (see text), values of $E_\mathrm{av}$ and fitted effective parameters $X_\mathrm{{fit}}$ (in cm$^{−1}$) for two lowest configurations of Nd$^{3+}$, Er$^{3+}$ and Eu$^{3+}$, as well as all the fitted parameters for Eu$^{3+}$. Here, for brevity, we use $F^k(fd)$ as $F^k(4f, 5d)$, $\zeta_{f}$ as $\zeta_{4f}$ and $\zeta_{d}$ as $\zeta_{5d}$.}
\begin{ruledtabular}
\begin{tabular}{c rrrrrrrr}
 Param. &  \multicolumn{4}{c}{$4f^w$} & \multicolumn{4}{c}{$4f^{w-1} 5d$} \\
   \cline{2-5} \cline{6-9}
  name & & & & & & & & \\
  & SF(Nd$^{3+}$) & SF(Er$^{3+}$) & SF(Eu$^{3+}$) & value (Eu$^{3+}$) & SF(Nd$^{3+}$) &  SF(Er$^{3+}$) & SF(Eu$^{3+}$) & value (Eu$^{3+}$) \\ 
  \hline
 $E_\mathrm{av}$ & 24898.0 & 35577.1 & & 65609.0 & 88430.0 & 133432.5 & & 137500.0\\
 $F^2(ff)$ & 0.738 & 0.754 & 0.781 & 88732.0 & 0.759 & 0.756 & 0.758 & 91269.9 \\
 $F^4(ff)$ & 0.825 & 0.919 & 0.950 & 67778.0 & 0.909 & 0.988 & 0.949 & 72029.8\\
 $F^6(ff)$ & 0.773 & 0.898 & 0.800 & 41060.7 & 0.870 & 0.798 & 0.834 & 45638.6 \\
 $\alpha$ & 19.1 & -0.2 & & 22.6 & 22.6 & 32.4 & &  22.6\\
 $\beta$  & -558.5 & -204.7 &  & -605.0 & -605.0  & -668.4 & &  -605.0 \\
 $\gamma$ & 1690.5 & 55.8 & & 292.2 & 292.2 & 1409.8 & & 292.2 \\
 $\zeta_{f}$ & 0.930 & 0.979 & 0.928 & 1313.6 & 0.947 & 0.995 & 0.971 & 1481.1 \\
 $\zeta_{d}$ &  & &  &  & 0.972 & 0.916 & 0.945 & 1290.0 \\
 $F^1(fd)$  &  & & &  & 1025.3 & 1370.6 & & 0 \\
 $F^2(fd)$  &  & & &  & 0.726 & 0.776 & 0.751 & 23046.8 \\
 $F^3(fd)$  &  & & &  & 111.5 & 2330.4 & & 0 \\
 $F^4(fd)$  &  & & &  & 1.128 & 1.124 & 1.126 & 16815.7 \\
 $G^1(fd)$  &  & & &  & 0.762 & 0.653 & 0.707 & 9113.2 \\
 $G^2(fd)$  &  & & &  & 2199 & 411.1 & & 0 \\
 $G^3(fd)$  &  & & &  & 1.005 & 0.838 & 0.922 & 10103.2 \\
 $G^4(fd)$  &  & & &  & 2016.0 &  0 & & 0 \\
 $G^5(fd)$  &  & & &  & 0.874 & 0.680 & 0.778 & 6603.8\\
\end{tabular}
\end{ruledtabular}
\end{table*}

The interpretation of Nd$^{3+}$ and Er$^{3+}$ spectra show that, because CI mixing is very low, a one-configuration approximation can safely be applied in both parities, which is done here. For Nd$^{3+}$, experimentally known levels are taken from the article of Wyart \textit{et al.}~\cite{wyart2007}. There are 41 levels for $4f^3$ configuration and 111 for $4f^2 5d$ configuration. For Er$^{3+}$, 38 experimental levels of the configuration $4f^{11}$ and 58 of $4f^{10} 5d$ are taken from Meftah \textit{et al.}~\cite{meftah2016}. For the $4f^6$ configuration of Eu$^{3+}$, the NIST database gives 12 levels \cite{NIST_ASD}.

Table \ref{tab:param_sf} shows a comparison of the final SFs (for \textit{ab initio} parameters) or the fitted values (for effective parameters), for the two lowest configurations of the above mentioned ions. It also shows the parameter values used in the Eu$^{3+}$ spectrum calculations of the next subsections. In the $4f^w$ configurations, the least-square fitting calculations, performed for each element, illustrates the regularities of SFs for $F^k$ and $\zeta_f$ parameters. Regarding effective parameters, the negative values of $\beta$ are usual, while the small values of $\alpha$ and $\gamma$ of Er$^{3+}$ are not. The regularities are also visible between $4f^2 5d$ and $4f^{10} 5d$ configurations of Nd$^{3+}$ and Er$^{3+}$ respectively. Therefore, we calculate our Eu$^{3+}$ parameters by multiplying the HFR values by the average SF obtained for Nd$^{3+}$ and Er$^{3+}$. The effective parameters are those obtained for Nd$^{3+}$, and the center-of-gravity energy of $4f^5 5d$ is calculated by assuming that the difference $E_\mathrm{av}(4f^{w-1}5d) - E_\mathrm{av}(4f^w)$ increases linearly with $w$.

Figure \ref{fig:energy_scheme} shows the levels computed with the parameters of Table \ref{tab:param_sf}, whose energies are between 0 and 100000~cm$^{-1}$, for the $4f^6$ and $4f^5 5d$ configurations. They will be analyzed in details in the next two paragraphs.

\subsection{\label{sub:eu_ground}Energy levels of the ground configuration $4f^{6}$}

\begin{table*}
\caption{\label{tab:Eu_ground}Comparison between the experimental, computed and other theory \cite{freidzon2018} values for the levels of $4f^6$ configuration of Eu$^{3+}$, with total angular momenta from $J=0$ to 6 and energies up to 30000~cm$^{-1}$, as well as first three LS-coupling eigenvectors with their percentages. All energy values are in cm$^{−1}$.}
\begin{ruledtabular}
\begin{tabular}{rrrrlrlrlr}
 Exp. & This work & Other theory & $J$  &  \multicolumn{6}{c} {First three eigenvectors and percentages }\\ \hline
 & & & & & & & & &\\
    0 &   -21 &     0 & 0 &  $^7$F  & 93.4 \% &  $^5$D1 &  3.5 \% &  $^5$D3 &  2.8 \%  \\
  370 &   357 &   380 & 1 &  $^7$F  & 94.7 \% &  $^5$D1 &  2.8 \% &  $^5$D3 &  2.2 \%  \\
 1040 &  1022 &  1040 & 2 &  $^7$F  & 96.3 \% &  $^5$D1 &  1.9 \% &  $^5$D3 &  1.4 \%  \\
 1890 &  1880 &  1880 & 3 &  $^7$F  & 97.4 \% &  $^5$D1 &  1.1 \% &  $^5$D3 &  0.7 \%  \\
 2860 &  2860 &  2830 & 4 &  $^7$F  & 97.9 \% &  $^5$F2 &  0.5 \% &  $^5$D1 &  0.4 \%  \\
 3910 &  3912 &  3860 & 5 &  $^7$F  & 97.6 \% &  $^5$G1 &  0.8 \% &  $^5$G3 &  0.8 \%  \\
 4940 &  4998 &  4970 & 6 &  $^7$F  & 96.4 \% &  $^5$G1 &  1.5 \% &  $^5$G3 &  1.5 \%  \\
17270 & 17257 & 17830 & 0 &  $^5$D3 & 45.4 \% &  $^5$D1 & 30.4 \% &  $^3$P6 &  6.7 \%  \\
19030 & 19015 & 19450 & 1 &  $^5$D3 & 50.6 \% &  $^5$D1 & 33.5 \% &  $^7$F  &  4.7 \%  \\
21510 & 21489 & 22140 & 2 &  $^5$D3 & 54.3 \% &  $^5$D1 & 36.1 \% &  $^7$F  &  2.9 \%  \\
24390 & 24360 & 25370 & 3 &  $^5$D3 & 55.2 \% &  $^5$D1 & 37.7 \% &  $^5$D2 &  2.0 \%  \\
      & 25257 &       & 6 &  $^5$L  & 88.7 \% &  $^3$K5 &  3.0 \% &  $^3$K1 &  2.2 \%  \\
      & 26314 &       & 2 &  $^5$G3 & 40.6 \% &  $^5$G1 & 36.0 \% &  $^5$G2 & 16.7 \%  \\
      & 26622 &       & 3 &  $^5$G3 & 37.8 \% &  $^5$G1 & 33.3 \% &  $^5$G2 & 16.4 \%  \\
      & 26814 &       & 4 &  $^5$G3 & 33.2 \% &  $^5$G1 & 28.5 \% &  $^5$G2 & 17.8 \%  \\
      & 26913 &       & 5 &  $^5$G3 & 30.2 \% &  $^5$G1 & 24.9 \% &  $^5$G2 & 20.2 \%  \\
      & 26926 &       & 6 &  $^5$G3 & 26.9 \% &  $^5$G2 & 22.7 \% &  $^5$G1 & 20.4 \%  \\
27640 & 27574 & 28960 & 4 &  $^5$D3 & 52.8 \% &  $^5$D1 & 37.6 \% &  $^5$F2 &  2.3 \%  \\

\end{tabular}
\end{ruledtabular}
\end{table*}

For the $4f^6$ configuration, values from the NIST database \cite{NIST_ASD} were taken as the experimentally known energy levels. Because the free ion has not been analyzed yet, the energies was determined by interpolation or extrapolation of known experimental values or by semi-empirical calculation \cite{martin1978}. Table \ref{tab:Eu_ground} shows a good agreement between these experimental values, our computed values and the theoretical values calculated by Freidzon and coworkers \cite{freidzon2018}. Our values are closer to the experimental ones in the $^5$D manifold. Note that a direct comparison with the article of Ogasawara and coworkers \cite{ogasawara2005} is difficult, as the authors do not give tables of energy levels for Eu$^{3+}$. In total, the $4f^6$ configuration contains 296 levels with $J$ values ranging from 0 to 12.

Table \ref{tab:Eu_ground} also illustrates that the ground-configuration levels are well described by the LS coupling scheme. Some levels are mainly characterized by a single term, like $^7$F or $^5$L, but others are shared between several terms with the same $L$ and $S$ quantum numbers, but different seniority numbers like $^5$D(1,2,3) or $^5$G(1,2,3), which are used to indicate that these are coming from different parent terms of $4f^5$ (see subsection \ref{sub:2nd}). The small deviations from LS coupling are due to the SO interaction, for example, a small $^5$D component in the $^7$F levels. The terms coupled by SO are such that $\Delta L=0,\pm 1$ and $\Delta S=0,\pm 1$ in agreement with Eq.~\eqref{eq:hso-g}.

Finally, Table \ref{tab:Eu_ground_manifold} contains the energy value and eigenvector of the manifolds with $S=2$ and 3, calculated by setting  to 0 the spin-orbit parameter $\zeta_f$ of Table \ref{tab:param_sf}. This information is necessary to build our third-order theory, see Eq.~\eqref{eq:0th-man-gr}. Note that the first excited manifold is a superposition of $^5$D3, $^5$D1 and $^5$D2 terms. But due to its strong importance in Eu$^{3+}$ spectroscopic studies, it will be denoted $^5$D in the rest of the paper.

\begin{table}
\caption{\label{tab:Eu_ground_manifold} Manifolds of quintet ($S=2$) and septet ($S=3$) multiplicities of $4f^6$ ground configuration of Eu$^{3+}$. All energy values are in cm$^{−1}$.}
\begin{ruledtabular}
\renewcommand{\arraystretch}{1.45} 
\begin{tabular}{rc}
Energy & Eigenvector \\ \hline
3895 & $\mid$$^7$F$\rangle$  \\
24561 & $\sqrt{0.576}$ $\mid$$^5$D3$\rangle$ - $\sqrt{0.406}$ $\mid$$^5$D1$\rangle$ - $\sqrt{0.019}$ $\mid$$^5$D2$\rangle$ \\
28212 & $\mid$$^5$L$\rangle$ \\
28268 & $\sqrt{0.408}$ $\mid$$^5$G3$\rangle$ - $\sqrt{0.328}$ $\mid$$^5$G1$\rangle$ - $\sqrt{0.264}$ $\mid$$^5$G2$\rangle$ \\
32821 & $\sqrt{0.652}$ $\mid$$^5$H1$\rangle$ - $\sqrt{0.348}$ $\mid$$^5$H2$\rangle$ \\
35683 & $\sqrt{0.970}$ $\mid$$^5$I2$\rangle$ - $\sqrt{0.03}$ $\mid$$^5$I1$\rangle$ \\
35822 & $\sqrt{0.755}$ $\mid$$^5$F2$\rangle$ - $\sqrt{0.245}$ $\mid$$^5$F1$\rangle$ \\
39329 & $\mid$$^5$K$\rangle$ \\
42446 & $\sqrt{0.732}$ $\mid$$^5$G2$\rangle$ - $\sqrt{0.159}$ $\mid$$^5$G1$\rangle$ - $\sqrt{0.109}$ $\mid$$^5$G3$\rangle$ \\
43892 & $\sqrt{0.803}$ $\mid$$^5$D3 $\rangle$ - $\sqrt{0.165}$ $\mid$$^5$D1$\rangle$ - $\sqrt{0.031}$ $\mid$$^5$D2$\rangle$ \\
45888 &  $\mid$$^5$P$\rangle$ \\
47553 & $\sqrt{0.652}$ $\mid$$^5$H2$\rangle$ - $\sqrt{0.348}$ $\mid$$^5$H1$\rangle$ \\
57251 &  $\mid$$^5$S$\rangle$ \\
62588 & $\sqrt{0.970}$ $\mid$$^5$I1$\rangle$ - $\sqrt{0.030}$ $\mid$$^5$I2$\rangle$ \\
64164 & $\sqrt{0.755}$ $\mid$$^5$F1$\rangle$ - $\sqrt{0.245}$ $\mid$$^5$F2$\rangle$ \\
75177 & $\sqrt{0.513}$ $\mid$$^5$G1$\rangle$ - $\sqrt{0.483}$ $\mid$$^5$G3$\rangle$ - $\sqrt{0.004}$ $\mid$$^5$G2$\rangle$ \\
76112 & $\sqrt{0.430}$ $\mid$$^5$D1$\rangle$ - $\sqrt{0.392}$ $\mid$$^5$D3$\rangle$ - $\sqrt{0.177}$ $\mid$$^5$D2$\rangle$ \\

\end{tabular}
\end{ruledtabular}
\end{table}

\subsection{\label{sub:eu_excited}Energy levels of the first excited configuration $4f^{5} 5d$}

\begin{table*}
\caption{\label{tab:Eu_excited} {First 20 energy levels for $4f^5 5d$ configuration of Eu$^{3+}$ with total angular momentum $J=0$, 1 and 2, as well as first three eigenvectors with their percentages. As an example, ($^6$H) $^7$H is used for brevity for $4f^5$($^6$H$^o$)$5d$ $^7$H$^o$, where the superscript ``o'' indicates odd parity. All energy values are in cm$^{−1}$.}}
\begin{ruledtabular}
\begin{tabular}{rrlrlrlr}
 Energy &  $J$  &  \multicolumn{6}{c} {First three eigenvectors and percentages} \\ \hline
  & & & & & & & \\
78744 & 2 & ($^6$H) $^7$H & 57.6 \% & ($^6$H) $^5$G & 14.1 \% & ($^6$F) $^7$H & 14.0 \%\\
79541 & 1 & ($^6$H) $^5$F & 52.9 \% & ($^6$H) $^7$G & 20.7 \% & ($^6$H) $^7$F & 10.1 \% \\
80396 & 2 & ($^6$H) $^5$F & 41.2 \% & ($^6$H) $^7$F & 23.7 \% & ($^6$H) $^7$G & 12.0 \%\\
81171 & 0 & ($^6$H) $^7$F & 91.5 \% & ($^4$G) $^5$D &  3.1 \% & ($^4$G) $^5$D &  2.0 \%\\
81493 & 1 & ($^6$H) $^7$F & 65.0 \% & ($^6$H) $^7$G & 23.1 \% & ($^6$F) $^7$G &  4.3 \%\\
82105 & 2 & ($^6$H) $^7$G & 48.6 \% & ($^6$H) $^7$F & 34.1 \% & ($^6$F) $^7$G &  9.5 \%\\
83096 & 1 & ($^6$H) $^7$G & 32.1 \% & ($^6$H) $^5$F & 26.6 \% & ($^6$H) $^7$F & 17.4 \%\\
83849 & 2 & ($^6$H) $^7$F & 31.6 \% & ($^6$H) $^5$F & 18.1 \% & ($^6$H) $^5$G & 14.7 \%\\
84398 & 1 & ($^6$F) $^7$G & 73.0 \% & ($^6$H) $^7$G & 18.5 \% & ($^6$F) $^5$F &  3.9 \%\\
84785 & 2 & ($^6$F) $^7$G & 75.1 \% & ($^6$H) $^7$G & 14.8 \% & ($^6$F) $^5$F &  2.4 \%\\
85060 & 2 & ($^6$F) $^7$H & 47.4 \% & ($^6$H) $^5$G & 18.3 \% & ($^6$H) $^5$F & 10.9 \%\\
86736 & 0 & ($^6$F) $^7$F & 81.5 \% & ($^6$F) $^5$D &  7.4 \% & ($^6$H) $^7$F &  2.6 \%\\
87056 & 1 & ($^6$F) $^7$F & 82.2 \% & ($^6$F) $^5$D &  6.7 \% & ($^6$H) $^7$F &  2.4 \%\\
87134 & 2 & ($^6$H) $^5$G & 37.2 \% & ($^6$H) $^7$H & 28.1 \% & ($^6$F) $^7$H & 18.1 \%\\
87679 & 2 & ($^6$F) $^7$F & 80.9 \% & ($^6$F) $^5$D &  5.9 \% & ($^6$P) $^7$F &  2.2 \%\\
89165 & 1 & ($^6$F) $^7$D & 84.2 \% & ($^6$F) $^5$P &  7.8 \% & ($^4$F) $^5$P &  1.8 \%\\
89220 & 2 & ($^6$F) $^7$P & 78.1 \% & ($^6$F) $^7$D &  6.0 \% & ($^6$F) $^5$P &  5.0 \%\\
90024 & 2 & ($^6$F) $^7$D & 81.8 \% & ($^6$F) $^7$P &  7.6 \% & ($^6$F) $^5$P &  2.0 \%\\
91979 & 0 & ($^6$F) $^5$D & 61.5 \% & ($^6$F) $^7$F &  7.9 \% & ($^6$P) $^5$D &  5.9 \%\\
93243 & 2 & ($^6$F) $^5$D & 47.0 \% & ($^6$F) $^5$G & 16.9 \% & ($^6$F) $^5$F &  5.6 \%\\

\end{tabular}
\end{ruledtabular}
\end{table*}

This subsection is devoted to the energy levels of the first excited configuration $4f^5 5d$. The parameters necessary for the calculations are given in Table \ref{tab:param_sf}. The $4f^5 5d$ configuration contains 1878 levels with $J$-values from 0 to 14, and according to our calculations, with energies from 74438 to 243060~cm$^{-1}$. The dominant eigenvector of the 74438-cm$^{-1}$ level is $4f^5(^6 \mathrm{H}^o) \,5d (^7 \mathrm{K}^o_4)$ with 93.8~\%. As examples, Table \ref{tab:Eu_excited} shows the 20 lowest energy levels with $J=0$, 1 and 2, along with their three dominant eigenvectors.

Table \ref{tab:Eu_excited} shows that the levels of the $4f^5 5d$ configuration do not possess a strongly dominant eigenvector (or a group of eigenvectors) characterized by the same $L$ and $S$ quantum numbers. This means that, unlike the ground configuration, see Table \ref{tab:Eu_ground}, the LS coupling scheme is not appropriate for the excited configuration. It can be shown that the $jj$ coupling scheme is not appropriate either, because the spin-orbit energy of the $5d$ electron is of the same order of magnitude as the electrostatic energy between $4f$ and $5d$ electrons. The eigenvectors are therefore written in pair coupling, \textit{i.e.}~linear combination of LS-coupling states.

In a given energy level, the $\overline{L}$ and $\overline{S}$ quantum numbers, which characterize the parent term of the $4f^5$ subshell, are common to the majority of the eigenvectors. With increasing energy, the levels mainly possess $^6$H$^o$, $^6$F$^o$ and $^6$P$^o$ characters; then come the quartet and doublet parent terms. Indeed the SO interaction within the $4f^5$ subshell is too small to significantly mix different $ \overline{L}$ and $\overline{S}$ of the $4f^5$ subshell. By contrast, the total $L$ and $S$ quantum numbers of the LS states differ at most by one unity. For example, we notice the pairs $^7$H-${}^5$G ($\Delta S = 1$ and $\Delta L = 1$), $^7$G-${}^7$F ($\Delta S = 0$ and $\Delta L = 1$) and $^5$F-${}^7$F ($\Delta S = 1$ and $\Delta L = 0$) for the level at 78744, 79541 and 80396~cm$^{-1}$ respectively. In consequence, the mixing between quintet and septet states of Eu$^{3+}$ is mainly due to the SO interaction of the $5d$ electron. That is why we ignore the influence of the $4f$ electrons to account for the Wybourne-Downer mechanism (see Subsection \ref{sub:3rd} and Eq.~\eqref{eq:hso-e}).

\subsection{\label{sub:eu_transition}Free-ion transitions between the two configurations}

\begin{figure}
\includegraphics[scale=0.47]{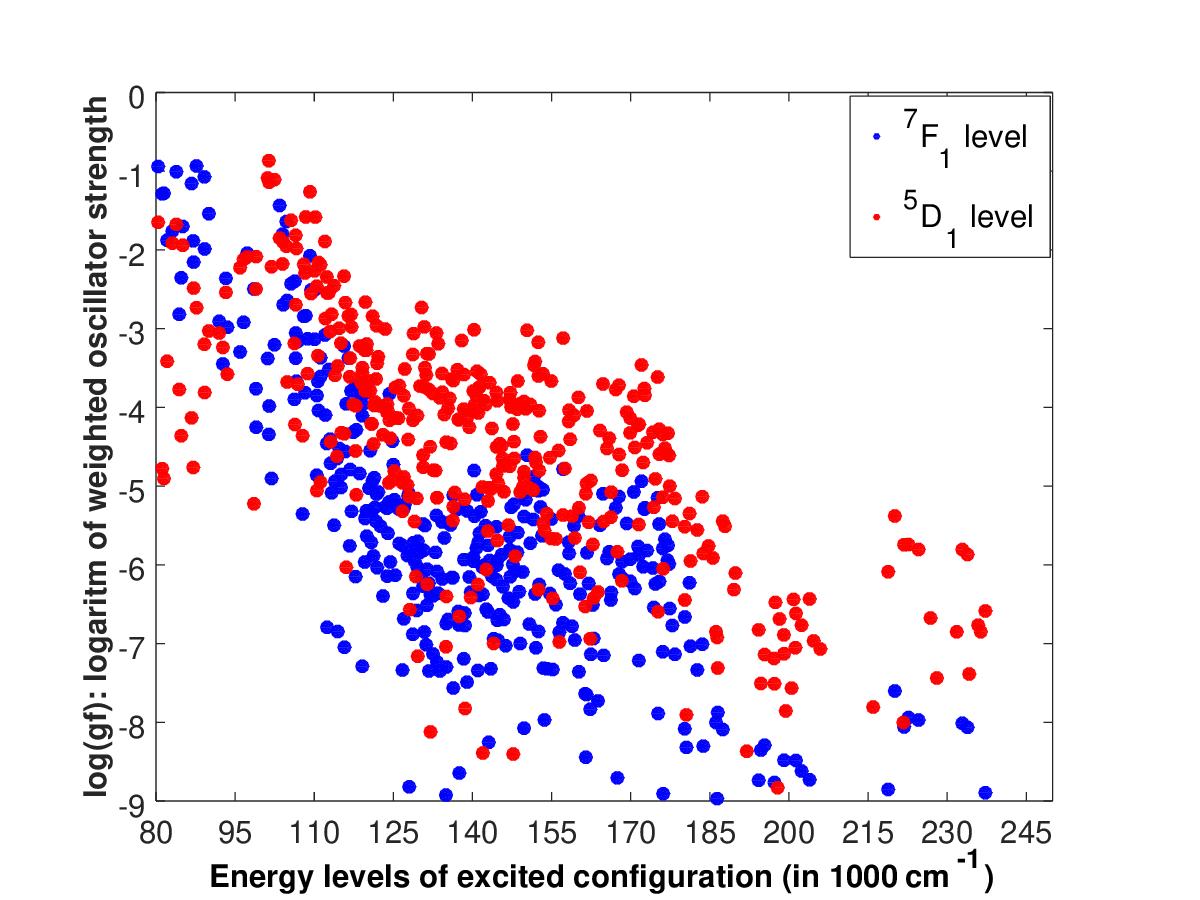}
\caption{\label{fig:log_gf} Logarithm of the weighted ED oscillator strengths, see Eq.~\eqref{eq:weighted_gf}, as functions of the energy of the excited-configuration levels, for transitions implying the $^7$F$_1$ (blue) and $^5$D$_1$ (red) levels of the ground configuration.}
\end{figure}

Equations \eqref{eq:sed-ext-1} and \eqref{eq:sed-ext-2} show that our f-f transition line strengths require the reduced multipole moments of some free-ion transitions which only occur between levels of the ground and excited configurations. In this subsection, we focus on the electric-dipole (ED) free-ion transitions ($k=1$), that are the most intense.

A widely used quantity for the discussion of spectral lines and transitions is the absorption oscillator strength $f_{12,\mathrm{ED}}$, which is related to the ED line strength $S_{\mathrm{ED}}$ through the expression
\begin{equation}
  f_{12,\mathrm{ED}} = \frac{2 m_e a_0^2 (E_2 - E_1)}{3\hbar^2 (2J_1+1)}
    \mathcal{S_{\mathrm{ED}}},
  \label{eq:f-abs}
\end{equation}
where 1 (2) denotes the lower (upper) levels of energy $E_1$ ($E_2$) and total angular momentum $J_1$ ($J_2$), $\hbar$ is the reduced Planck constant, $m_e$ the electron mass, $a_0 = 4\pi\epsilon_0\hbar^2 / m_e e^2$ the Bohr radius, $\epsilon_0$ the vacuum permitivity and $e$ the electron charge. In Eq.~\eqref{eq:f-abs}, the ED line strength is in atomic units (units of $e^2 a_0^2$). Because the oscillator strength for stimulated emission is defined as $f_{21} = - \frac{2J_1+1}{2J_2+1} f_{12}$, the so-called weighted oscillator strength
\begin{equation}
  gf_\mathrm{ED} = (2J_1+1)f_{12,\mathrm{ED}}
   = -(2J_2+1) f_{21,\mathrm{ED}}
\label{eq:weighted_gf}
\end{equation}
does not depend on the nature of the transition. For ED free-ion transitions, the line strength of Eq.~\eqref{eq:f-abs} is the square of the reduced ED matrix element, $\mathcal{S_{\mathrm{ED}}} = |\langle \Psi_1 \Vert P^{(1)} \Vert \Psi_2 \rangle |^2$. In the rest of the article, we will focus on the absorption oscillator strengths, and so will drop the {}``12" subscripts.

Figure \ref{fig:log_gf} shows the dependence of the logarithm of the weighted oscillator strengths given by Eq.~\eqref{eq:weighted_gf} on the energy of the excited-configuration level, for transitions involving two levels of the ground configuration. It shows that the energy band with strong transitions is rather narrow and lies in the range of 80000–100000 cm$^{−1}$, while for larger excited-level energies, the values of $\log(gf)$ for the level $^7$F$_1$ (blue dots) decrease faster than those for $^5$D$_1$ (red dots). Indeed, the total spin $S$ of $4f^5 5d$ levels tends to decrease with energy (see Table \ref{tab:Eu_excited}), the coupling with levels of the $4f^6\,{}^7$F manifold drops faster than the coupling with levels of the quintet manifolds. Therefore, in the framework of the JO theory, the excited-configuration energy $E_{n'\ell'}$ appearing in the denominators of the line strengths, see Eqs.~\eqref{eq:sed-ext-1} and \eqref{eq:sed-ext-2}, is not the center-of-gravity energy of the excited-configuration, but rather the strong-coupling window between 80000 and 100000~cm$^{−1}$: in practice, we take $E_{n'\ell'} = 90000$~cm$^{−1}$.

In addition to the free-ion ED reduced matrix elements, the JO theory requires those for $k=3$ (octupole) and $k=5$, which depend on the radial transition integral $\langle n' l' | r^k | n l \rangle = \int^{\infty}_0 dr P_{n' l'}(r) r^k P_{n l}(r)$,~where $n l=4 f$ and $n' l'=5 d$. We have calculated those integrals with a home-made Octave code, reading the HFR radial wave functions $P_{4f}$ and $P_{5d}$ computed by Cowan's code RCN. We obtain 1.130629$\,a_0$, -3.221348$\,a_0^3$ and 21.727152$\,a_0^5$ for $k=1$, $k=3$ and $k=5$, respectively, while the $k=1$ value calculated by Cowan is 1.130618$\,a_0$.

\subsection{\label{sub:ff_transitions}f-f transitions in Eu$^{3+}$-doped solids}

Now that we have all the necessary information about the free-ion spectrum, in this subsection, we aim to benchmark our model with experimental data. To that end, we have chosen the thorough investigation of Babu \textit{et al.}~\cite{babu2000}, who measured absorption oscillator strengths and interpreted them with the standard JO theory. Their study deals with transitions within the ground manifold $^7$F and between the ground and first excited manifold $^5$D for Eu$^{3+}$-doped lithium fluoroborate glass. In the latter case, the transitions involve a change in spin, well known to challenge the standard JO theory.

\subsubsection{Description of our calculations}

We have written a FORTRAN program which firstly reads the energies and four leading eigenvectors of the ground-configuration free-ion levels (see Table \ref{tab:Eu_ground}) and manifolds (see Table \ref{tab:Eu_ground_manifold}). Then, the code performs a linear least-square fitting of experimental line strengths $S_\mathrm{exp}$ and the ED part of the theoretical ones $S_\mathrm{ED}$ given by Eqs.~\eqref{eq:sed-ext-1} and \eqref{eq:sed-ext-2}, with the free adjustable parameters
\begin{equation}
  X_k = \sum_{q=-k}^{+k} \frac{|A_{kq}|^2}{2k+1}
  \label{eq:xk}
\end{equation}
for $k=1$, 3 and 5, which describes the electrostatic environment at the ion position. During the least-square step, we seek to minimize the standard deviation on line strengths
\begin{equation}
  \sigma = \left[ \frac{ \sum_{i=1}^{N_\mathrm{tr}} 
    \left( S_{\mathrm{exp},i}-S_{\mathrm{ED},i} \right)^2}
    {N_\mathrm{tr}-N_\mathrm{par} }  \right]^{\frac{1}{2}},
  \label{eq:sigma}
\end{equation}
where $N_\mathrm{tr}$ is the number of transitions included in the calculation and $N_\mathrm{par}=3$ is the number of adjustable parameters. The experimental line strengths are extracted from the absorption oscillator strengths $f_\mathrm{exp}$ by inverted Eq.~\eqref{eq:f-abs},
\begin{equation}
  S_\mathrm{exp} = \frac{3(2J_1+1) \hbar^2}{2 m_e a_0^2 (E_2 - E_1)}
    \frac{n_r}{\chi_\mathrm{ED}} f_\mathrm{exp}
  \label{eq:s-exp}
\end{equation}
where $n_r$ is the host refractive index, and $\chi_\mathrm{ED} = (n_r^2+2)/9$ the local-field correction in the virtual-cavity model (see for example Ref.\cite{aubret2018}). In contrast with the free-ion case, Eq.~\eqref{eq:s-exp} takes into account the host material through its refractive index $n_r$; for lithium fluoroborate, $n_r=1.57$ is assumed wavelength-independent. Note that our code can also apply the fitting procedure to Einstein $A$ coefficients, as the latter are transformed in line strengths.

After the fitting, using these optimal $X_k$ parameters, we can predict line strengths, oscillator strengths and Einstein $A$ coefficients, for other transitions. Of course, that procedure only involves transitions with a predominant ED character; magnetic-dipole (MD) transitions like $^5$D$_0 \leftrightarrow {}^7$F$_1$ and $^5$D$_1 \leftrightarrow {}^7$F$_0$ are therefore excluded from the fit. For them, the MD line strength $\mathcal{S}_\mathrm{MD}$, oscillator strengths and Einstein coefficients can be calculated from the free-ion eigenvectors (see Table \ref{tab:Eu_ground}) \cite{dodson2012}.

\subsubsection{Results of the least-square fitting}

\begin{table}
\caption{\label{tab:judd_ofelt}{Transition labels, experimental oscillator strengths ($\times$ 10 $^{-6}$) \cite{babu2000} and ratios between theoretical and experimental line strength for the standard Judd-Ofelt theory ($r_0$), as well as for our second-order ($r_1$) and third-order ($r_2$) corrections of our theory. The last lines present the absolute and relative standard deviations for each model (see text).}}
\begin{ruledtabular}
\begin{tabular}{ccccc}
Tr. label & Exp. o.s. & $r_0$ & $r_1$ & $r_2$ \\ \hline
 & & & & \\
$^5$D$_4 \leftrightarrow {}^7$F$_0$ & 0.489 & 1.14 & 0.07 & 0.80 \\
$^5$G$_2 \leftrightarrow {}^7$F$_0$ & 0.523 & 0.82 & 0.59 & 0.28 \\ 
$^5$L$_6 \leftrightarrow {}^7$F$_0$ & 3.338 & 0.38 & 0.31 & 0.29 \\
$^5$L$_6 \leftrightarrow {}^7$F$_1$ & 1.383 & 0.17 & 0.33 & 0.29 \\
$^5$D$_3 \leftrightarrow {}^7$F$_1$ & 0.302 & 0.87 & 0.12 & 1.10 \\
$^5$D$_2 \leftrightarrow {}^7$F$_0$ & 0.333 & 1.41 & 1.66 & 0.88 \\
$^5$D$_1 \leftrightarrow {}^7$F$_1$ & 0.450 & 0.99 & 1.00  & 1.00 \\
$^7$F$_6 \leftrightarrow {}^7$F$_0$ & 1.232 & 1.83 & 1.81 & 1.83  \\
$^7$F$_6 \leftrightarrow {}^7$F$_1$ & 1.983 & 0.93 & 0.95  & 0.94\\ \hline 
& & & & \\
$\sigma$ & & 0.266 & 0.252 & 0.258 \\ 
$\sigma/\mathcal{S}_\mathrm{max}$ & & 8.71~\% & 8.24~\% & 8.45~\% 
\end{tabular}
\end{ruledtabular}
\end{table}

\begin{figure}
\includegraphics[scale=0.65]{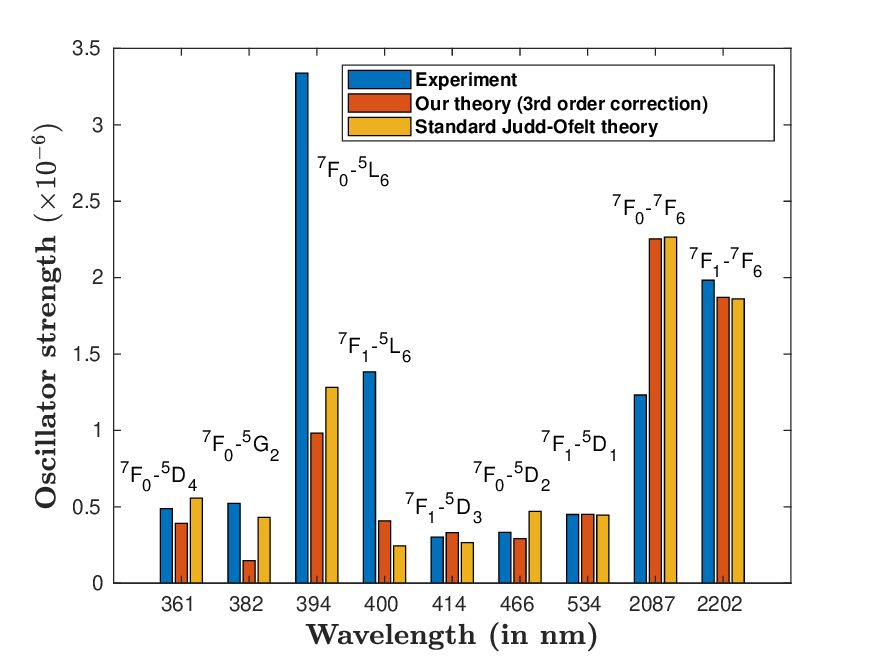}
\caption{\label{fig:histogram_log_gf} Comparison between experimental \cite{babu2000} and theoretical oscillator strengths of absorption, plotted as function of the transition wavelength (not at scale). The transitions are labeled with the LS-term quantum numbers of the Eu$^{3+}$ free ion.}
\end{figure}

We have included 9 out of the 14 transitions measured with the so-called L6BE glass in Table 3 of Ref.~\cite{babu2000}. We have excluded three predominant MD transitions, $^5$D$_0 \leftrightarrow {}^7$F$_1$, $^5$D$_1 \leftrightarrow {}^7$F$_0$ and $^5$D$_2 \leftrightarrow {}^7$F$_1$, as well as the $^5$G$_4 \leftrightarrow {}^7$F$_0$ and $^5$D$_0 \leftrightarrow {}^7$F$_0$ for which we observe large deviations between theory and experiment. They are probably due to the fact that the $^5$D$_0$ and $^5$G$_4$ are further from LS coupling than the other levels. In particular, the four leading components represent 88.4 and 86.9~\% of the total eigenvectors respectively.

Table \ref{tab:judd_ofelt} shows the results of our least-square calculations with the second- and third-order theory, in comparison with the standard JO theory used in Ref.~\cite{babu2000}. For each transition, the table contains the experimental values of the oscillator strength ($\times$10$^{-6} $) \cite{babu2000} and the ratios $r_n$ between the theoretical and experimental oscillator strengths, where $r_0$ is the ratio for the standard Judd-Ofelt theory, and the $r_1$ and $r_2$ are the ratios, respectively, for second and third order corrections of theory (see subsection \ref{sub:2nd} and \ref{sub:3rd}). For each model, we present the absolute $\sigma$ and relative standard deviations, taken by dividing Eq.~\eqref{eq:sigma} by the largest experimental line strength $\mathcal{S}_\mathrm{max} = 3.057 \times 10^{-4}$ for the $^7$F$_6 \leftrightarrow {}^7$F$_1$ transition. Figure \ref{fig:histogram_log_gf} gives a visual insight into the results of Table \ref{tab:judd_ofelt}, with histograms of the experimental absorption oscillator strengths, and those resulting from the standard JO theory and our third-order correction, plotted as functions of the transition wavelength.

Globally, the three methods have similar performances. That shows that the SO interaction in the excited configuration has little effect, since it is included in the third-order correction and not in the second-order one. Our third-order corrections better describes transitions between $^7$F and $^5$D manifolds. However, it predicts the smallest oscillator strength for $^5$G$_2 \leftrightarrow {}^7$F$_0$, owing to the proximity between the $^5$G3 and $^5$H1 manifolds (see Table \ref{tab:Eu_ground_manifold}), which puts into question the use of SO interaction as a perturbation. On the other hand, the second-order correction fails to describe the $^5$D$_4 \leftrightarrow {}^7$F$_0$ transition. The three methods tend to underestimate the oscillator strengths for high-energy transitions, where the refractive index $n_r$ is larger than 1.57.

\begin{table}
\caption{\label{tab:fit_param} Fitted parameters from the absorption oscillator strengths of Ref.~\cite{babu2000}. The second column gives standard JO parameters $\Omega_{2,4,6}$; the third and fourth ones give $X_k$ obtained with Eqs.~\eqref{eq:sed-ext-1} and \eqref{eq:sed-ext-2}, respectively.}
\begin{ruledtabular}
\begin{tabular}{crrr}
  & Std. JO & 2nd-order & 3rd-order \\ \cline{2-4}
 $k$ & $\Omega_{k+1}^{\phantom{I^I}}$ ($10^{-20}\,$cm$^2$) & $X_k$ (a.u.) & $X_k$ (a.u.) \\
\hline
 1 & $\phantom{{}^{I^I}}$17.93 & $9.424 \times 10^{-7}$ & $1.441 \times 10^{-6}$ \\
 3 & 11.92 & $2.330 \times 10^{-5}$ & $9.916 \times 10^{-6}$ \\
 5 &  2.13 & $7.187 \times 10^{-8}$ & $6.406 \times 10^{-8}$
\end{tabular}
\end{ruledtabular}
\end{table}

The final fitting parameters are given in Table \ref{tab:fit_param} for the standard JO calculation of Ref.~\cite{babu2000} (see Set B of Table 4), as well as our second-order correction \eqref{eq:sed-ext-1} and third-order correction \eqref{eq:sed-ext-2}. The orders of magnitude of the $X_k$ are the same for the two corrections. The parameter $X_3$ are the largest, then the $X_1$ are roughly one order of magnitude smaller than the $X_3$, and the $X_5$ are roughly two orders of magnitude smaller than $X_3$. It is hard to make direct comparisons with the standard JO parameters given in Table 4 of Ref.~\cite{babu2000} (data set B); but we see that that the $\Omega_6$ parameter, responsible of the $^7$F$_6 \leftrightarrow {}^7$F$_{0,1}$ and $^5$L$_6 \leftrightarrow {}^7$F$_{0,1}$ transitions just like $X_5$ is respectively 9 and 6 times smaller than $\Omega_2$ and $\Omega_4$. To give more insight values of the parameter, we notice that the quantities $\sqrt{X_k} \times \langle n\ell |r^k| n'\ell' \rangle$ is the order-of-magnitude energy of the ion-field interaction: in the third-order correction, they are respectively equal to 298, 2226 and 1207~cm$^{-1}$ for $k=1$, 3 and 5.

\subsubsection{Transitions with a MD character}

Now that we have the $X_k$ parameters, we can calculate oscillator strengths for transitions not present in the fit. In particular, we can predict the percentage of ED and MD characters for the transitions having both characters \cite{freed1941, kunz1980, taminiau2012, chacon2020}, assuming that the total oscillator strength is equal to the sum $f_\mathrm{ED} + f_\mathrm{MD}$. The ED part can be calculated by inverting Eq.~\eqref{eq:s-exp} and replacing the subscripts {}``exp'' by {}``ED'', while the MD part reads \cite{dodson2012}
\begin{equation}
  f_\mathrm{MD} = \frac{2 m_e a_0^2 (E_2 - E_1)} {3(2J_1+1) \hbar^2}
   n_r \mathcal{S}_\mathrm{MD}
  \label{eq:f-md}
\end{equation}
where the MD line strength is written in units of $e^2 a_0^2$ \cite{cowan1981}
\begin{equation}
  \mathcal{S}_\mathrm{MD} = \frac{\alpha^2}{4}
    \left| \left\langle \Psi_1 \right\Vert \mathbf{L} + g_s\mathbf{S} \left\Vert \Psi_2 \right\rangle \right|^2
  \label{eq:s-md}
\end{equation}
with $\alpha$ the fine-structure constant and $g_s$ the electronic-spin g-factor. Because the orbital $\mathbf{L}$ and spin $\mathbf{S}$ angular momenta are even-parity tensors of rank one, MD transitions can occur in free space or in solids, between levels of the same configuration and with $\Delta J\le 1$ except $(J_1,J_2)=(0,0)$.

\begin{table}
\caption{\label{tab:ed-part} Experimental absorption oscillator strengths, as well as ED and MD theoretical ones, for transitions having both an ED and a MD character (numbers in brackets are the powers of 10).}
\begin{ruledtabular}
\begin{tabular}{clll}
Tr. label & $f_\mathrm{exp}$ & $f_\mathrm{th,ED}$ & $f_\mathrm{th,MD}$ \\ \hline
 & & & \\
$^5$D$_1 \leftrightarrow {}^7$F$_0$ &  7.8(-8) &  2.74(-9) & 2.67(-8)  \\
$^5$D$_0 \leftrightarrow {}^7$F$_1$ &  5.7(-8) & 7.70(-10) & 3.93(-8)  \\
$^5$D$_1 \leftrightarrow {}^7$F$_1$ & 4.50(-7) &  4.50(-7) & 1.92(-11) \\
$^5$D$_2 \leftrightarrow {}^7$F$_1$ & 2.48(-7) &  1.74(-7) & 4.96(-9)  \\
\end{tabular}
\end{ruledtabular}
\end{table}

Table \ref{tab:ed-part} presents experimental and theoretical absorption oscillator strengths for the transitions having in principle an ED and a MD character. The MD oscillator strengths are calculated by multiplying the free-ion one computed with Cowan's code by the host refractive index $n_r$, see Eq.~\eqref{eq:f-md}. The table clearly shows that the $^5$D$_1 \leftrightarrow {}^7$F$_1$ transition is purely electric (at least 99.9~\%), hence its inclusion in the fit. The $^5$D$_2 \leftrightarrow {}^7$F$_1$ is also mainly electric, but to a lesser extent, roughly at 95~\%. The two others are mostly magnetic, but the experimental and theoretical MD oscillator strengths significantly differ from each other. Still, the ED character looks larger for the $^5$D$_1 \leftrightarrow {}^7$F$_0$ transition (1-2~\%) than for the $^5$D$_0 \leftrightarrow {}^7$F$_1$ one (4-9~\%).

\subsubsection{The $^5$D$_0 \leftrightarrow {}^7$F$_0$ transition}

Since the $^5$D$_0 \leftrightarrow {}^7$F$_0$ transition is forbidden by the selection rules of the standard JO model, it has attracted a lot of attention (see Ref.~\cite{binnemans2015} and references therein), in order to understand its origin. Even though it is not forbidden in our model, we had to exclude it in the fit, because of a strong discrepancy between the experimental and our computed oscillator strength. With our optimal parameters $X_k$, we obtain an oscillator strength $1.25 \times 10^{-7}$, that is 7.8 times larger than the experimental value. In this paragraph, we investigate in closer details the possible origin of that discrepancy and how to reduce it.

Firstly, as mentioned in Subsection \ref{sub:3rd}, the sums in Eqs.~\eqref{eq:d12-4} and \eqref{eq:sed-ext-2} involves quintet and septet manifolds of the ground configuration. But a closer look at the eigenvectors shows that the $^5$D$_0$ level contains 6.7~\% of $^3$P6 character, see Table \ref{tab:Eu_ground}, as well as 5.1~\% of $^3$P3, while $^7$F$_0$ contains 0.1~\% of $^3$P6 character. These small components are likely to contribute to the transition amplitude, and so they need to be accounted for in a future work, through a complete description of the free-ion eigenvectors.

The selection rules associated with Eq.~\eqref{eq:d12-4} show merely the terms with $k=1$ of the CF potential can induce a transition of the kind $(J_1,J_2) = (0,0)$. This result seems consistent because: (i) those terms are stronger in sites with low symmetries, and (ii) observing the $^5$D$_0 \leftrightarrow {}^7$F$_0$ transition is an indication of $C_{nv}$, $C_n$ or $C_s$ point groups at the ion site \cite{nieuwpoort1966, binnemans1996, binnemans1997}.

Another frequently invoked mechanism to explain the $^5$D$_0 \leftrightarrow {}^7$F$_0$ transition is $J$-mixing \cite{tanaka1994, kushida2002, kushida2003}, especially between levels of the lowest manifold $^7$F. However, because this mixing is limited to 10~\%, it cannot explain the strongest 0-0 transitions listed in Ref.~\cite{chen2005}. Charge-transfer states are also likely to play a role in the 0-0 transition, especially in hosts with oxygen-compensating sites around by which the CF tends to be strongly deformed \cite{karbowiak2000}. However, those two mechanisms are not present in our model.

\subsubsection{Radiative lifetime of the $^5$D$_0$ level}

In addition to absorption oscillator strengths, our model also makes it possible to calculate the ED Einstein coefficient for the spontaneous emission from level 2 to 1,
\begin{equation}
  A_{\textrm{ED}} = \frac{e^2 a_0^2 (E_2 - E_1)^3}
    {3\pi \epsilon_0 \hbar^4 c^3 (2J_2+1)}
    n_r \chi_{\mathrm{ED}} \mathcal{S}_{\mathrm{ED}},
  \label{eq:a-em}
\end{equation}
where $\mathcal{S}_\mathrm{ED}$ is given by Eq.~\eqref{eq:sed-ext-2}. We can also compute the MD Einstein coefficients $A_{\textrm{MD}}$, by multiplying the free-ion value calculated with Cowan by $n_r^3$; namely
\begin{equation}
  A_\mathrm{MD} = \frac{e^2 a_0^2 (E_2 - E_1)^3}
    {3\pi \epsilon_0 \hbar^4 c^3 (2J_2+1)} n_r^3
   \mathcal{S}_\mathrm{MD}
\end{equation}
where $\mathcal{S}_\mathrm{MD}$ is given by Eq.~\eqref{eq:s-md}.

From them, we can deduce the radiative lifetime $\tau$ of a given level. In particular for the $^5$D$_0$ level, it reads
\begin{equation}
  \tau ({}^5\mathrm{D}_0) = \left( \sum_{J=0}^{6}
      A_{\textrm{ED}}({}^5\mathrm{D}_0,{}^7\mathrm{F}_J)
    + A_{\textrm{MD}}({}^5\mathrm{D}_0,{}^7\mathrm{F}_1)
  \right)^{-1} .
\end{equation}
Transitions $^5D_0 \leftrightarrow {}^7F_J$, where $J = 1$ to 6, are not included in our fit, and so are considered as additional transitions, for which our program calculated line strengths and Einstein coefficients. For the transition $^5D_0 \leftrightarrow {}^7F_1$, the total Einstein coefficient is the sum of the electric and the magnetic parts, calculated using Cowan code. The latter is found to be $A_{\textrm{MD}}({}^5\mathrm{D}_0,{}^7\mathrm{F}_1)$ = 53.44 s$^{-1}$. The sum of Einstein coefficients for all other transitions, including the electric part of transition $^5D_0 \leftrightarrow {}^7F_1$, is 500.529 s$^{-1}$. That sum includes the transition $^5D_0 \leftrightarrow {}^7F_0$, whose Einstein coefficient \eqref{eq:a-em} is calculated using the line strength deduced from the experimental oscillator strength following Eq.~\eqref{eq:s-exp}. This yields the very small value of 0.029 s$^{-1}$. The resulting radiative lifetime is $\tau({}^5\mathrm{D}_0) = 1805~\mu$s, which is close to the experimental value of 1920~$\mu$s reported in Ref.~\cite{babu2000}. In principle, the relaxation limiting the lifetime is due to radiative as well as nonradiative processes; however the latter are expected to be unlikely for the $^5$D$_0$ level \cite{rabouw2016}, due to the large gap between the $^5$D$_0$ and $^7$F$_6$ levels, see Table \ref{tab:Eu_ground}.

\section{\label{sec:conclusion} Conclusion}

In this article, we have developed an extension of the Judd-Ofelt model enabling to calculate intensities of absorption and emission transitions for Ln$^{3+}$-doped solids. In our model, the properties of the Ln$^{3+}$ impurity are fixed parameters calculated with free-ion spectroscopy, while the crystal-field ones are adjusted by least-square fitting. In particular, the line strengths, oscillator strengths and Einstein coefficients are functions of three least-square fitted crystal-field parameters.

We have benchmarked our model with a detailed spectroscopic study of europium-doped lithium borate glasses. Not only our model allows for giving a simple physical insight into the transitions which are not described by the standard Judd-Ofelt theory, but it also reproduces measured oscillator strengths with a similar accuracy to the standard theory \cite{babu2000}. Moreover, we demonstrate that the spin-changing transitions in Eu$^{3+}$ mainly result from the spin-orbit mixing within the ground electronic configuration, even if its levels are well described by the LS-coupling scheme.

In consequence, our model may be improved in the future, by taking into account all the eigenvector components of the free-ion levels, while the four leading ones are taken into account in the current study. We expect this improvement to give more a precise calculation of the ${}^7$F$_0 \leftrightarrow {}^5$D$_0$ intensity. We also plan to account for the wavelength-dependence of the refractive index of the host material. Finally, the fact of separating the dopant and crystal-field parameters opens the possibility to interpret transitions between individual crystal-field levels or involving polarized light, which is especially relevant for nanometer-scale host materials.

In contrast, spectroscopic studies of free Ln$^{3+}$ ions indicate that configuration-interaction mixing, between the configurations $4f^w$ and $4f^{w-1}6p$ on the one hand, $4f^{w-1}5d$ and $4f^{w-1}6s$ on the other hand, does not have a strong role in the energy spectrum \cite{wyart2007, meftah2016}, and so shall not be included in our model. However, in the case of Er$^{3+}$ \cite{arab2019}, the lowest core-excited configuration of opposite parity compared to the ground one, $5p^5 4f^3 5d$ starts at 182000~cm$^{-1}$. Assuming a similar order of magnitude for the $5p^5 4f^6 5d$ configuration of Eu$^{3+}$, and taking the relativistic Hartree-Fock value of the radial integral $\langle 5p |r| 5d \rangle = 1.62\,a_0$, we can expect the excitation of the $5p$ core electrons toward the $5d$ orbital, to have a sizeable effect on the crystal-field coupling to opposite-parity configurations.

\section*{\label{sec:acknowledgements}Acknowledgements}

We would like to thank Reinaldo Chac{\'o}n, G{\'e}rard Colas des Francs and Aymeric Leray for fruitful discussions. We acknowledge support from the NeoDip project (ANR-19-CE30-0018-01 from ``Agence Nationale de la Recherche''). M.L. also acknowledges the financial support of {}``R{\'e}gion Bourgogne Franche Comt{\'e}'' under the projet 2018Y.07063 {}``Th{\'e}CUP''. Calculations have been performed using HPC resources from DNUM CCUB (Centre de Calcul de l'Universit\'e de Bourgogne).

\appendix

\section{Useful relationships}
\label{sec:app-theory}

In the appendix, the LS-coupling basis functions of the ground and excited configurations are respectively written $|\alpha LSJ \rangle$ and $|\overline{\alpha} \overline{L} \overline{S}, \, L'S'J' \rangle$. The reduced matrix element of the electric-multipole operator $P_q^{(k)}$ are given by \cite{cowan1981}
\begin{widetext}
\begin{align}
  \left\langle \overline{\alpha} \overline{L} \overline{S}, \, L'S'J' \right\Vert P^{(k)} \left\Vert \alpha LSJ \right\rangle
  & = \left(-1\right)^{S+J+\overline{L}+k} \sqrt{w\left(2J+1\right) \left(2J'+1\right) \left(2L+1\right) \left(2L'+1\right) \left(2\ell+1\right) \left(2\ell'+1\right)} \nonumber \\
  & \times (n\ell^{w-1}\, \overline{\alpha} \overline{L} \overline{S} \rrbracket n\ell^{w}\, \alpha LS) \left\{ \begin{array}{ccc}
    L & S & J\\
    J' & k & L'
  \end{array}\right\} \left\{ \begin{array}{ccc}
    \ell & \overline{L} & L\\
    L' & k & \ell'
  \end{array}\right\} \left(\begin{array}{ccc}
    \ell' & k & \ell\\
    0 & 0 & 0
  \end{array}\right)
  \left\langle n'\ell' \right| r^{k} \left| n\ell \right\rangle ,
  \label{eq:pq-red}
\end{align}
where $(n\ell^{w-1}\, \overline{\alpha} \overline{L} \overline{S} \rrbracket n\ell^{w}\, \alpha LS)$ is a coefficient of fractional parentage introduced by Racah \cite{racah1943}.
The matrix element of the spin-orbit operator within the ground configuration is
\begin{align}
  & \left\langle \alpha_{1} L_{1} S_{1} J_{1} \right| H_{\mathrm{SO}} \left| \alpha_{2} L_{2} S_{2} J_{2} \right\rangle \nonumber \\
  = & \,\delta_{J_{1}J_{2}}  \left(-1\right)^ {L_{2}+S_{1}+J_{1}} \, w\zeta_{n\ell}\, \sqrt{\ell\left(\ell+1\right) \left(2\ell+1\right) s\left(s+1\right) \left(2s+1\right) \left(2L_{1}+1\right) \left(2L_{2}+1\right)} \nonumber \\
  \times & \sqrt{\left(2S_{1}+1\right) \left(2S_{2}+1\right)} \sum_{\overline{\alpha} \overline{L} \overline{S}} \left(-1\right)^ {\ell+s+\overline{L}+\overline{S}+L_{1}+S_{1}} \left\{ \begin{array}{ccc}
    \ell & \ell & 1\\
    L_{1} & L_{2} & \overline{L}
  \end{array}\right\} \left\{ \begin{array}{ccc}
    s & s & 1\\
    S_{1} & S_{2} & \overline{S}
  \end{array}\right\} \nonumber \\
  \times & (n\ell^{w-1}\, \overline{\alpha} \overline{L} \overline{S} \rrbracket n\ell^{w}\, \alpha_{1} L_{1} S_{1}) \times (n\ell^{w-1}\, \overline{\alpha} \overline{L} \overline{S} \rrbracket n\ell^{w}\, \alpha_{2}L_{2}S_{2})
  \label{eq:hso-g}
\end{align}
In the excited configuration, we assume that the off-diagonal matrix elements are only due to the outermost $n'\ell'=5d$ electron,
\begin{align}
  & \left\langle  \overline{\alpha} \overline{L} \overline{S},\,L'_1 S'_1 J'_1 \right|H_{\mathrm{SO}} \left| \overline{\alpha} \overline{L} \overline{S},\, L'_2 S'_2 J'_2 \right\rangle \nonumber \\
  = & \,\delta_{J'_{1}J'_{2}}  \left(-1\right) ^ {\ell'+s+\overline{L}+\overline{S}+L'_{1}+L'_{2}+2S'_{1}+J'_{1}} \, \zeta_{n'\ell'}\, \sqrt{\ell'\left(\ell'+1\right) \left(2\ell'+1\right) s\left(s+1\right) \left(2s+1\right)} \nonumber \\
  \times & \sqrt{\left(2L'_{1}+1\right) \left(2L'_{2}+1\right) \left(2S'_{1}+1\right) \left(2S'_{2}+1\right)} 
  \sixj{L'_1}{S'_1}{J'_1}{S'_2}{L'_2}{1} 
  \sixj{\ell'}{\ell'}{1}{L'_1}{L'_2}{\overline{L}}
  \sixj{s}{s}{1}{S'_1}{S'_2}{\overline{S}}
  \label{eq:hso-e}
\end{align}
\end{widetext}
The second-order correction on the eigenvector $|\Psi_i^2 \rangle$ is
\begin{align}
  \left| \Psi_i^2 \right\rangle & 
   = \sum_{t,u\neq i} \left| \Psi_t^0 \right\rangle 
   \frac{V_{tu}V_{ui}}{\Delta_{it}\Delta_{iu}}
   - \sum_{t\neq i} \left| \Psi_t^0 \right\rangle 
   \frac{V_{ti}V_{ii}}{\Delta_{it}^2}
  \nonumber \\
  & - \frac{1}{2} \left| \Psi_i^0 \right\rangle \sum_{t\neq i} 
   \frac{V_{it}V_{ti}}{\Delta_{it}^2}
  \label{eq:2nd-eigvct}
\end{align}
where $V_{it} = \langle \Psi_i^0 |V| \Psi_t^0 \rangle$ and $\Delta_{it} = E_i - E_t$.



%

\end{document}